\documentclass[lettersize,journal]{IEEEtran}
\usepackage{amsmath,amsfonts,amssymb,mathtools}
\usepackage{algorithm}
\usepackage{algpseudocode} 
\usepackage{array}
\usepackage[caption=false,font=footnotesize]{subfig}
\usepackage{textcomp}
\usepackage{stfloats}
\usepackage{url}
\usepackage{verbatim}
\usepackage{graphicx}
\usepackage{cite}
\usepackage{balance}

\usepackage{xcolor}
\usepackage{comment}

\usepackage{multirow}
\usepackage{hyperref}
\usepackage{balance}
\usepackage[table]{xcolor}
\definecolor{minshade}{RGB}{230,240,255} 
\definecolor{maxshade}{RGB}{255,230,230} 

\begin{document}

\title{
Trajectory-Aware Air-to-Ground Channel Characterization for Low-Altitude UAVs Using MaMIMO Measurements}

\author{Abdul Saboor,~\IEEEmembership{Student Member,~IEEE}, 
        Zhuangzhuang Cui,~\IEEEmembership{Member,~IEEE}, 
        Achiel Colpaert,~\IEEEmembership{Member,~IEEE}, 
        Evgenii Vinogradov,~\IEEEmembership{Senior Member,~IEEE}, 
        Wout Joseph,~\IEEEmembership{Senior Member,~IEEE}
        and Sofie Pollin,~\IEEEmembership{Senior Member,~IEEE}%
\thanks{Abdul Saboor, Zhuangzhuang Cui, Achiel Colpaert, and Sofie Pollin are with WaveCoRE, Department of Electrical Engineering (ESAT), KU Leuven, 3001 Leuven, Belgium (e-mail: \{firstname.lastname\}@kuleuven.be). \textit{(Corresponding author: Zhuangzhuang Cui)}}%
\thanks{Achiel Colpaert, Zhuangzhuang Cui, and Sofie Pollin are also with imec, Kapeldreef 75, 3001 Leuven, Belgium.} %
\thanks{Wout Joseph is with INTEC-WAVES-IMEC, UGent, Ghent, Belgium (e-mail: wout.Joseph@ugent.be).}
\thanks{Evgenii Vinogradov is with the NaNoNetworking Center in Catalonia (N3Cat), Universitat Polit\`ecnica de Catalunya, Barcelona, Spain (e-mail: evgenii.vinogradov@upc.edu).}}

\markboth{IEEE Transactions on Vehicular Technology,~Vol.~xx, No.~xx, August~2025}%
{Saboor \MakeLowercase{\textit{et al.}}: Measurement-Based, Multi-Dimensional, Trajectory-Aware Channel Characterization for Low-Altitude UAV Communications}


\maketitle

\begin{abstract}

This paper presents a comprehensive measurement-based trajectory-aware characterization of low-altitude Air-to-Ground (A2G) channels in a suburban environment. A 64-element Massive Multi-Input Multi-Output (MaMIMO) array was used to capture channels for three trajectories of an Uncrewed Aerial Vehicle (UAV), including two horizontal zig-zag flights at fixed altitudes and one vertical ascent, chosen to emulate AUE operations and to induce controlled azimuth and elevation sweeps for analyzing geometry-dependent propagation dynamics. We examine large-scale power variations and their correlation with geometric features, such as elevation, azimuth, and 3D distance, followed by an analysis of fading behavior through distribution fitting and Rician K-factor estimation. Furthermore, temporal non-stationarity is quantified using the Correlation Matrix Distance (CMD), and angular stationarity spans are utilized to demonstrate how channel characteristics change with the movement of the UAV. We also analyze Spectral Efficiency (SE) in relation to K-factor and Root Mean Square (RMS) delay spread, highlighting their combined influence on link performance. The results show that the elevation angle is the strongest predictor of the received power, with a correlation of more than 0.77 for each trajectory, while the Nakagami model best fits the small-scale fading. The K-factor increases from approximately 5 dB at low altitudes to over 15 dB at higher elevations, indicating stronger LoS dominance. Non-stationarity patterns are highly trajectory- and geometry-dependent, with azimuth most affected in horizontal flights and elevation during vertical flight. These findings offer valuable insights for modeling and improving UAV communication channels in 6G Non-Terrestrial Networks (NTNs). 
\end{abstract}


\section{Introduction}

\IEEEPARstart{A}{lthough} 5G has significantly improved data rates, latency, and network capacity, communication resources remain largely centered on terrestrial networks \cite{10927643}. In contrast, 6G envisions extending connectivity well beyond the constraints of ground-based infrastructure \cite{siddiky2025comprehensive}. To achieve this, researchers are exploring networks in the vertical dimension by integrating Non-Terrestrial Networks (NTNs) in existing terrestrial networks. In the future, 6G will integrate NTNs, consisting of satellites, high- and low-altitude platforms, and ground nodes, all coming together to form a unified three-dimensional (3D) communication framework \cite{11010845, 10793277, guidotti2024role}.  

Within NTNs, Uncrewed Aerial Vehicles (UAVs) are gaining significant attention in expanding the connectivity and capacity of the existing terrestrial networks due to their rapid mobility, fast deployment, and high availability of Line-of-Sight links \cite{OJCS, mohsan2022towards, Blacksea}. These features enable UAVs, acting as an Aerial Base Station (ABS), to assist terrestrial infrastructure in scenarios of disrupted coverage or uneven user distribution, making them ideal for on-demand coverage, time-sensitive missions, and emerging 6G applications such as Intelligent Transportation Systems (ITS) \cite{Eucap, 10793277, gryech2024systematic}.

In the context of ITS, the role of UAVs is not limited to ABS. Instead, they can also operate as Aerial User Equipment (AUE) using existing and future cellular infrastructures to support different applications. These applications include passenger transport (e.g., flying taxis), flying cargo and parcel delivery, surveillance, and environmental monitoring \cite{betti2024uav, naveen2024unlocking}. Due to these advantages, Morgan Stanley projects that the AUE market will reach up to 11\% of the projected global Gross Domestic Product (GDP) by 2050 \cite{morganstanley2021_uam_tam}.

Characterizing the Air-to-Ground (A2G) wireless channel for AUE is critical for optimizing connectivity in NTN-enabled 6G systems. Generally, UAVs as AUE have frequent LoS access to multiple Ground Base Stations (GBSs); however, they also suffer from challenges, including rapid changes in geometry and different angular spreads compared to ground-based users. Consequently, A2G channels become highly non-stationary due to constantly changing 3D geometry with a strong influence from elevation and azimuth angles. These factors directly affect signal quality, delay spread, and data rates, making accurate A2G channel characterization vital for better network design and mobility management.

\subsection{Related Work}

Numerous studies have explored system performance, trajectory planning, and protocol development based on large-scale channel modeling \cite{agrawal2021performance, wang2021learning, saboor2025cash, zhang2021energy}. However, many rely on simplified models or assumptions that fail to represent real operating conditions. Hence, there remains a clear need for high-resolution, measurement-based research to accurately capture the unique propagation characteristics of UAV links in realistic environments.

A2G channel characterization for UAVs has been an active research topic in both ABS and AUE contexts~\cite{cui2022cluster, cui2020wideband, khawaja2017uav, tu2019low, lv2023narrowband, AWPL, saboor2025empirical}. Early measurement campaigns focused primarily on large-scale fading and probability of LoS ($P_{\mathrm{LoS}}$) estimation for urban and suburban environments \cite{tu2019low, lv2023narrowband, AWPL}. For example, Path Loss Exponents (PLEs) are reported in a range from 2.0–2.7 in suburban and 4.5–4.8 in campus environments \cite{tu2019low, lv2023narrowband}. In contrast, a Path Loss ($PL$) model is presented for vehicular and pedestrian users in urban environments using a geometric $P_{\mathrm{LoS}}$ approach \cite{AWPL}. These studies revealed the strong dependence of $P_{\mathrm{LoS}}$ on elevation angle and building density, but often relied on simplified geometry-based models and sparse empirical validation.

Unlike ground vehicles, AUEs experience rapid changes in both azimuth and elevation angles due to their 3D mobility \cite{eskandari2022model}. Therefore, several works have investigated A2G channels in suburban environments \cite{matolak2017air, xiao2025measurements}, analyzing the combined effects of mobility, height, and Multipath Components (MPCs) on metrics such as Root Mean Square (RMS) delay spread. These studies demonstrated that mobile AUE links experience rapid geometry changes, leading to strong non-stationarity in frequency, spatial, and temporal domains.

The advent of Massive MIMO (MaMIMO) and 3D beamforming has driven a shift toward more comprehensive channel characterizations that account for spatial, temporal, and angular statistics. Colpaert et al.  \cite{colpaertMimo} presented a 3D MaMIMO-UAV channel measurement campaign employing a large antenna array at GBS to study Power Delay Profiles (PDPs) and spatial stationarity. Their findings showed strong elevation-angle dependencies in the non-stationary. However, the authors considered straight-line trajectories, which often overlook the complex mobility patterns and non-stationary conditions present in practical AUE scenarios, such as surveillance, target tracking, or search-and-rescue missions, where zig-zag or irregular flight paths are more common.

To further explore A2G channel stationarity, several studies have proposed advanced Geometry-based Stochastic Models (GBSMs) for non-stationary UAV channels \cite{bai2022non, hua2025ultra, liu2021novel, bian20213d, bai2022non2}. Bai et al. \cite{bai2022non} developed a Space–Time–Frequency (STF) model for 6G mmWave MaMIMO UAV links using a correlated cluster birth–death algorithm, later extending it to irregular 3D trajectories. Other efforts investigated a maritime UAV channel model with sea-surface reflection effects \cite{liu2021novel}, and an ultra-wideband UAV–ground channel study focusing on frequency non-stationarity \cite{hua2025ultra}. While theoretically rich, these works are mainly based on simulations or Ray-Tracing (RT), and hence, they lack extensive validation through real-world measurement campaigns, which can limit their reliability in practical AUE scenarios. Lastly, Bain et al. \cite{bian20213d} proposed a wideband MIMO UAV–ground channel model that removes the constraint of the straight trajectory by combining an aeronautic Smooth Turn Random Mobility Model (STRMM) with a concentric cylinder geometry. Using simulations, the authors highlighted that UAV trajectories greatly influence channel statistics and increase temporal non-stationarity.

In summary, the current literature has the following gaps: 1) a scarcity of multi-domain measurement campaigns that jointly analyze delay-domain metrics and temporal stationarity under realistic UAV mobility; 2) limited empirical studies linking measured channel metrics to system-level performance; 3) limited attention to UAVs as AUE rather than ABS; 4) an over-reliance on simplistic linear flight paths, which under-represent angular diversity; and 5) a predominance of simulation-based studies with limited real-world measurement validation.  

To address these gaps, this work presents a high-resolution measurement campaign featuring multiple zig-zag flight trajectories, two at fixed altitudes and one with gradual ascent, covering heights up to 59 m (about twice the average building height) to capture a broad range of azimuth and elevation angles at 2.61 GHz. The main contributions of this paper are:

\begin{itemize}
    \item We conducted a field experiment with three UAV trajectories, including two zig-zag flights at fixed altitudes and one vertical ascent, covering the full 3D suburban space and yielding over $1.4\times10^5$ channel samples per trajectory~\cite{MTNAEG_2025}. The dataset enables reproducibility and serves as a basis for future AUE channel studies. The analysis reveals that elevation is the strongest predictor of received power, azimuth dominates fading in vertical flight, and small-scale fading follows a Nakagami distribution.
    \item We modeled the K-factor as a function of AUE height, observing a log-linear increase validated by both horizontal trajectories, offering a basis for height-dependent AUE link design.
    \item We extended stationarity analysis beyond spatial distance to include frequency, azimuth, and elevation, demonstrating that zig-zag trajectories reveal richer non-stationary behavior and that distance or frequency alone cannot distinguish vertical from horizontal flights.
    \item We linked channel characteristics to system performance using Spectral Efficiency (SE), which shows a moderate positive correlation with the K-factor and a weak negative correlation with the RMS delay spread, indicating that reduced multipath dispersion and stronger dominant components generally enhance performance. 
\end{itemize}

The rest of the paper is organized as follows: Section~\ref{sec2} describes the system model and channel analysis methodology. Section~\ref{sec3} details the measurement setup and AUE trajectories. Section~\ref{sec4} reports geometry-driven channel behavior and Section~\ref{statSec} discusses trajectory-driven stationarity results with corresponding performance evaluation. Finally, Section~\ref{sec5} concludes the paper and outlines future research directions.

\begin{table}[!t]
\centering
\caption{List of Symbols}
\label{SymbolsTab}
\renewcommand{\arraystretch}{1.2}
\begin{tabular}{c l}
\hline
\textbf{Symbol} & \textbf{Description} \\
\hline
$a$ , $b$ & Logarithmic fit parameters for K-factor model \\
$A_{\text{az}}(t_i)$ & Azimuth stationarity span at time $t_i$ \\
$A_{\text{el}}(t_i)$ & Elevation stationarity span at time $t_i$ \\
$a_{\mathrm{SSF}}[n]$ & Small-scale fading envelope at sample $n$ \\
$B$ & Local frequency range for correlation matrix \\
$B_{\text{coh}}(t_i)$ & Coherence bandwidth at time $t_i$ \\
$d_{\mathrm{2D}}[n]$, $d_{\mathrm{3D}}[n]$ & Horizontal and 3D distance at sample $n$ \\
$d_{\text{CMD}}(t_i,t_j)$ & Correlation Matrix Distance between $t_i$ and $t_j$ \\
$D_{\text{stat}}(t_i)$ & Temporal stationarity distance at time $t_i$ \\
$\Delta d[n]$ & Traveled distance between samples \\
$\Delta f$ & Frequency offset \\
$\Delta f_{+}(t_i)$, $\Delta f_{-}(t_i)$ & +ve/-ve frequency offset for coherence bandwidth \\
$\Delta \mathbf{p}_n$ & Relative position vector from GBS to AUE \\
$\eta(t_i)$ & Spectral efficiency at time $t_i$ \\
$F$ & Number of OFDM subcarriers \\
$g(t)$ & Narrowband channel model \\
$G(t)$ & Instantaneous received power \\
$G_a$ & Mean received power \\
$G_v$ & RMS fluctuation of received power \\
$\gamma$ & CMD threshold for stationarity \\
$h$ & UAV/AUE altitude (height) \\
$h_m(t,\tau)$ & Channel impulse response at antenna $m$ \\
$\mathbf{h}^{f}(t_k,\Delta f)$ & Frequency-dependent receive vector \\
$\mathbf{H}[n]$ & Channel matrix at time index $n$ \\
$K_{\mathrm{dB}}$ & Rician K-factor in dB \\
$L$ & Physical window length \\
$L_{\min}$ & Minimum stationarity distance \\
$M$ & Number of receive antenna elements \\
$N$ & Number of time instants/samples \\
$N_\tau$ & Number of delay taps \\
$N_w$ & Analysis window size in samples \\
$p(t)$ & Instantaneous power at time $t$ \\
$p_{\mathrm{LS}}[n]$ & Large-scale fading component \\
$\mathbf{p}_n$ & UAV position vector at sample $n$ \\
$\mathbf{p}_{\mathrm{BS}}$ & GBS position vector \\
$P_h(t_i,\tau)$ & Averaged Power Delay Profile \\
$P_m(t_i)$ & Mean power at time $t_i$ \\
$\varphi[n]$ & Azimuth angle at sample $n$ \\
$\rho$ & Correlation coefficient \\
$\mathbf{R}_a(t_i)$ & Receive correlation matrix at time $t_i$ \\
$R_f(t_i,\Delta f)$ & Frequency-domain correlation function \\
$S_\tau(t_i)$ & RMS delay spread at time $t_i$ \\
$t_{\min}, t_{\max}$ & Stationarity region boundaries \\
$T_m(t_i)$ & Mean delay at time $t_i$ \\
$\theta[n]$ & Elevation angle at sample $n$ \\
$V$ & Deterministic LoS component \\
$v(t)$ & Complex Gaussian multipath component \\
$|V|^2$ & LoS power component \\
$\bar{v}$ & Mean UAV velocity \\
$W$ & Temporal window size \\
$W_{\mathrm{LS}}$ & Large-scale fading window size \\
$\xi$ & Signal-to-noise ratio \\
\hline
\end{tabular}
\end{table}

\section{System Model and Analysis Methodology}
\label{sec2}

This section presents the mathematical modeling and signal processing steps used to characterize the wireless A2G channel experienced by a low-altitude AUE. The analysis integrates three-dimensional geometry, statistical fading, and non-stationary channel behavior. For clarity, all key system parameters used in this study are listed in Table \ref{SymbolsTab}.

\subsection{Channel Model and Geometry}
We consider a time-varying A2G MIMO channel, where a UAV-mounted single-antenna transmitter (AUE) communicates with a fixed ground-based MaMIMO receiver (GBS) at UAV altitudes up to 59~m. The received channel is sampled over $N$ time instants, with each snapshot represented by a complex-valued matrix:

\begin{equation}
\label{eqqq1}
\mathbf{H}[n] \in \mathbb{C}^{M \times F}, \quad n = 1, 2, \dots, N,
\end{equation}
\noindent
where $M$ is the number of receive antenna elements and $F$ is the number of Orthogonal Frequency-Division Multiplexing (OFDM) subcarriers. Each element of the matrix, $H_{m,f}[n]$, represents the complex baseband channel coefficient at time index $n$, corresponding to antenna element $m \in \{1, \dots, M\}$ and subcarrier index $f \in \{1, \dots, F\}$. This coefficient captures the combined effects of large-scale path loss, small-scale fading, and frequency selectivity for each antenna–subcarrier pair. Let the UAV position at discrete time index $n$ be given by the Cartesian coordinate vector $\mathbf{p}_n$:
\begin{equation}
\mathbf{p}_n = [x_n, y_n, z_n]^\top \in \mathbb{R}^3,
\end{equation}
and the position of the GBS is denoted as
\begin{equation}
\mathbf{p}_{\mathrm{BS}} = [x_{\mathrm{BS}}, y_{\mathrm{BS}}, z_{\mathrm{BS}}]^\top.
\end{equation}

The relative position vector from the GBS to the AUE is then
\begin{equation}
\Delta \mathbf{p}_n = \mathbf{p}_n - \mathbf{p}_{\mathrm{BS}} = [\Delta x_n, \Delta y_n, \Delta z_n]^\top.
\end{equation}

Based on this relative geometry, the following propagation-related quantities are defined:
\begin{align}
d_{\mathrm{2D}}[n] &= \sqrt{(\Delta x_n)^2 + (\Delta y_n)^2}, \\
d_{\mathrm{3D}}[n] &= \sqrt{(\Delta x_n)^2 + (\Delta y_n)^2 + (\Delta z_n)^2}, \\
\theta[n] &= \arctan \left( \frac{\Delta z_n}{d_{\mathrm{2D}}[n]} \right), \\
\varphi[n] &= \arctan \left( \frac{\Delta y_n}{\Delta x_n} \right).
\end{align}

\noindent
Here, $d_{\mathrm{2D}}[n]$ and $d_{\mathrm{3D}}[n]$ are the horizontal and 3D distances between the AUE and the GBS, respectively. Similarly, $\theta[n]$ and $\varphi[n]$ denote the elevation and azimuth angles. This spatial modeling provides the geometric basis for characterizing the channel in the delay, fading, and angular domains.

\subsection{Channel Stationarity}

UAV channels are inherently non-stationary due to continuous geometry changes. However, within limited time or frequency spans, they can be treated as Wide-Sense Stationary (WSS) if the mean power and correlation remain nearly constant~\cite{willink2008wide}. Stationarity is typically described in two domains:

\begin{itemize}
    \item \textbf{Frequency Stationarity:} The channel is locally stationary within a limited frequency span where its statistical properties remain nearly constant~\cite{cheng2022channel}.
    
    \item \textbf{Temporal Stationarity:} The UAV channel can be treated as stationary over a short time or traveled distance where its large-scale statistics remain nearly constant. Beyond this quasi-stationary region, UAV movement causes noticeable changes in the channel properties due to evolving geometry and multipath conditions.
\end{itemize}

Overall, frequency stationarity describes how the channel behaves across different frequencies, while temporal stationarity represents how it evolves along the UAV's path. Together, both stationarities define the temporal–frequency regions over which the channel can be treated as locally Wide-Sense Stationary.

\vspace{1em}
\subsubsection{\textbf{Frequency Stationarity}}
Frequency stationarity describes how the channel statistics vary across frequency. Each snapshot index $t_i$ corresponds to one UAV position (or time instant) from the measured frequency-domain channel matrix $\mathbf{H}[n]$ in~\eqref{eqqq1}. To analyze this behavior, the normalized frequency-domain correlation function $R_f(t_i,\Delta f)$ is computed using~\cite{bultitude2002estimating}:
\begin{equation}
R_f(t_i,\Delta f) = 
\frac{\mathbb{E}_f\!\left\{H(t_i,f) H^*(t_i,f+\Delta f)\right\}}
{\mathbb{E}_f\!\left\{|H(t_i,f)|^2\right\}}.
\label{eq:freq_corr}
\end{equation}

The coherence bandwidth $B_{\text{coh}}(t_i)$ quantifies the frequency range over which subcarriers experience highly correlated fading and, consequently, similar channel statistics. It is defined as the frequency spacing where the correlation magnitude first drops below $1/e$~\cite{he2015characterization}. For positive and negative frequency offsets, this can be expressed as
\begin{align}
\Delta f_{+}(t_i) &= \arg\!\max_{\Delta f>0}
\Big(|R_f(t_i,\Delta f)| = \tfrac{1}{e}\Big), \label{eq:df_pos}\\[4pt]
\Delta f_{-}(t_i) &= \arg\!\min_{\Delta f<0}
\Big(|R_f(t_i,\Delta f)| = \tfrac{1}{e}\Big). \label{eq:df_neg}
\end{align}
Finally, the coherence bandwidth is obtained as half the absolute difference between these frequency offsets:
\begin{equation}
B_{\text{coh}}(t_i) = \tfrac{1}{2}\big[\Delta f_{+}(t_i) - \Delta f_{-}(t_i)\big].
\label{eq:coh_bw}
\end{equation}

\vspace{0.3em}
\noindent In short, the channel is guaranteed to be frequency-stationary over the coherence bandwidth. The stationarity bandwidth can be obtained by combining several adjacent coherence regions exhibiting unchanged channel statistics.

\vspace{1em}
\subsubsection{\textbf{Temporal Stationarity}}
\label{cmdsubsec}

Temporal stationarity describes how long the channel statistics remain stable as the AUE moves. Over short distances, these properties remain nearly constant, defining a stationary region with distance $D_{\text{stat}}(t_i)$. To quantify this, the receive correlation matrix $\mathbf{R}_a(t_i)$ is computed within a small temporal window of $W$ samples and a local frequency range $B$ corresponding to the coherence bandwidth $B_{\text{coh}}(t_i)$. It is defined as

\begin{equation}
\mathbf{R}_a(t_i)
= \frac{1}{B W}
\sum_{\Delta f=-B/2}^{B/2}
\sum_{k=i}^{i+W-1}
\mathbf{h}^{f}(t_k,\Delta f)\,
\mathbf{h}^{f}(t_k,\Delta f)^{H},
\label{eq:Rx_corr}
\end{equation}
where $\mathbf{h}^{f}(t_k,\Delta f)$ is the frequency-dependent receive vector of $M$ antennas at time index $t_k$.  
By averaging over both $W$ and a frequency range $B \approx B_{\text{coh}}(t_i)$, this formulation ensures that $\mathbf{R}_a(t_i)$ captures only the locally stationary channel behavior.

The similarity between two channel states at positions $t_i$ and $t_j$ is quantified using the
\emph{Correlation Matrix Distance} (CMD)~\cite{herdin2005correlation}:
\begin{equation}
d_{\text{CMD}}(t_i,t_j) =
1 -
\frac{\mathrm{Tr}\!\left\{
\mathbf{R}_a(t_i)\,\mathbf{R}_a(t_j)
\right\}}
{\|\mathbf{R}_a(t_i)\|_F \,\|\mathbf{R}_a(t_j)\|_F},
\label{eq:CMD}
\end{equation}
where $\mathrm{Tr}\{\cdot\}$ denotes the trace operator and $\|\cdot\|_F$ is the Frobenius norm.  
A smaller $d_{\text{CMD}}$ indicates that the spatial channel structure remains highly similar between the two positions.

A region is considered temporally stationary (or quasi-stationary)  when the CMD value remains below a predefined threshold $\gamma$, typically chosen as 0.20~\cite{he2015characterization}:
\begin{equation}
d_{\text{CMD}}(t_i,t_j) \leq \gamma.
\label{eq:CMD_threshold}
\end{equation}

The corresponding temporal limits of this quasi-stationary region are determined by finding the first and last positions where the CMD remains below the threshold $\gamma$, expressed by
\begin{align}
t_{\min} &= \arg\!\max_{0 \le j \le i-1} \; d_{\text{CMD}}(t_i, t_j) \ge \gamma, \label{eq:tmin}\\[3pt]
t_{\max} &= \arg\!\min_{i+1 \le j \le T-W} \; d_{\text{CMD}}(t_i, t_j) \ge \gamma. \label{eq:tmax}
\end{align}
Using these boundaries, the temporal stationarity distance is given by
\begin{equation}
D_{\text{stat}}(t_i)
= [t_{\max}(t_i) - t_{\min}(t_i)] \, \bar{v},
\label{eq:Dstat_formal}
\end{equation}
where $\bar{v}$ denotes the mean UAV velocity or average traveled distance per sample. By projecting these positions onto the azimuth and elevation domains, the corresponding stationary angular spans can be derived as
\begin{align}
A_{\text{az}}(t_i) &= \varphi_{\max}(t_i) - \varphi_{\min}(t_i), \label{eq:Aaz}\\[3pt]
A_{\text{el}}(t_i) &= \theta_{\max}(t_i) - \theta_{\min}(t_i). \label{eq:Ael}
\end{align}

\noindent
Together, $D_{\text{stat}}(t_i)$, $A_{\text{az}}(t_i)$, and $A_{\text{el}}(t_i)$ describe how long and over what angular range the channel can be regarded as locally stationary.  
These parameters are later used to define the averaging window for estimating local metrics such as the K-factor, RMS delay spread, and spectral efficiency.

\subsection{Small-Scale Envelope Extraction}
To analyze the fading characteristics of the measured A2G channel, we first extract the received envelope and decompose it into large-scale and small-scale components. Let $H(t,m,f)$ be the measured channel at snapshot $t$, receive antenna $m$, and subcarrier $f$.
We form the instantaneous power averaged over antennas and subcarriers
\begin{equation}
p(t) = \frac{1}{MF}\sum_{m=1}^{M}\sum_{f=1}^{F}\!\bigl|H(t,m,f)\bigr|^{2}.
\label{eq:pt_power}
\end{equation}

The large-scale fading component is extracted by smoothing the instantaneous received power over a spatial window of length $L = 60\lambda$, which provides more conservative averaging than the standard 20-40$\lambda$ range~\cite{lee2006estimate} to ensure adequate suppression of small-scale fading. Since the measurements are sample-based, this physical window length is mapped to the discrete sample domain. Let $\Delta d[n] = d_{\mathrm{3D}}[n] - d_{\mathrm{3D}}[n-1]$ denote the AUE’s traveled distance between two consecutive samples, and 
$\mathbb{E}[\Delta d[n]]$ be the average step size. The number of samples corresponding to the physical window $W_{\mathrm{LS}}$ can be then approximated as:

\begin{equation}
    W_{\mathrm{LS}} = \frac{L}{\mathbb{E}[\Delta d[n]]}.
    \label{eq:WLS}
\end{equation}

The large-scale fading is obtained by applying a moving average filter over $W_{\mathrm{LS}}$ samples as
\begin{equation}
    p_{\mathrm{LS}}[n] = 
    \frac{1}{W_{\mathrm{LS}}}
    \sum_{i=n-\lfloor W_{\mathrm{LS}}/2 \rfloor}^{\,n+\lfloor W_{\mathrm{LS}}/2 \rfloor} 
    \tilde{p}[i],
    \label{eq:LSF}
\end{equation}
where $\tilde{p}[n]$ is the instantaneous power obtained after removing short-term spikes and outliers. Finally, the Small-Scale fading (SSF) envelope is then extracted by normalizing the instantaneous power with its large-scale component:
\begin{equation}
    a_{\mathrm{SSF}}[n] = 
    \sqrt{\frac{\tilde{p}[n]}{p_{\mathrm{LS}}[n]}}.
    \label{eq:SSF}
\end{equation}
This normalization removes path loss and shadowing, producing a small-scale unit-mean-power amplitude suitable for distribution fitting and K-factor estimation.

\subsection{Windowed Estimation of Local Channel Metrics}


In this section, we restrict all local (i.e., window-based) metrics to short spatial segments where the channel can be considered approximately WSS. 

Let $\{D_{\text{stat}}(t_s)\}_{s}$ denote the set of stationarity distances extracted along the trajectory. After removing outliers caused by noise or abrupt local decorrelations, the minimum representative stationarity distance is defined as
\begin{equation}
L_{\min} = \min\{D_{\text{stat}}(t_s)\},
\end{equation}

This value of $L_{\min}$ reflects a physically meaningful lower bound on the quasi-stationary region, excluding outlier points.

The minimum stationarity length $L_{\min}$ is mapped to discrete samples using the mean step size $\Delta d[n]$, 
giving an analysis window 
$N_w = \mathrm{round}\!\left(\frac{L_{\min}}{\Delta d[n]}\right)$. 
A hop size of $\lfloor N_w / 2 \rfloor$ ensures a 50\% overlap between successive windows.

Finally, we calculate all local parameters, such as K-factor, RMS delay spread, and SE, within this minimum stationarity window. The key idea of using $L_{\min}$ is to ensure that each estimate is obtained over a region where the channel is approximately WSS, avoiding the influence of large-scale variations like path loss and shadowing on small-scale metrics. The 50\% overlap improves statistical reliability while maintaining spatial resolution along the trajectory.

\vspace{1em}

\subsubsection{\textbf{Rician K-Factor}}
To evaluate communication stability, we compute the Rician K-factor, representing the ratio of the LoS component to the MPCs in the channel. Within each locally stationary window $L_{\min}$, the narrowband channel can be modeled as
\begin{equation}
g(t) = V + v(t),
\label{eq:g_model}
\end{equation}
where $V$ is the deterministic component and $v(t)$ is a zero-mean complex Gaussian variable representing multipath fading. For K-factor estimation, we follow the moment-based method proposed by Greenstein et al.~\cite{Greenstein769521}, which derives K from the first and second-order moments of the received signal power. The resulting instantaneous received power is
\begin{equation}
G(t) = |g(t)|^2 = |V|^2 + |v(t)|^2 + 2\,\mathrm{Re}\!\{Vv^*(t)\}.
\end{equation}
Averaging over the local window yields the mean power and its RMS fluctuation:
\begin{align}
G_a &= \mathbb{E}[G(t)] = |V|^2 + \sigma^2, \label{eq:Ga}\\
G_v &= \sqrt{\mathbb{E}\!\big[(G(t) - G_a)^2\big]} 
     = \sqrt{4|V|^2\sigma^2 + 2\sigma^4}, \label{eq:Gv}
\end{align}
where $\sigma^2 = \mathbb{E}[|v(t)|^2]$ is the diffuse power. Using \eqref{eq:Ga}–\eqref{eq:Gv}, $|V|^2$ and $\sigma^2$ are obtained as
\begin{align}
|V|^2 &= \tfrac{1}{2}\!\left(G_a + \sqrt{G_a^2 - G_v^2}\right),\\
\sigma^2 &= G_a - |V|^2.
\end{align}
Finally, the Rician K-factor is computed as
\begin{equation}
K = \frac{|V|^2}{\sigma^2}, 
\qquad K_{\mathrm{dB}} = 10\log_{10}(K).
\label{eq:K_final}
\end{equation}

A key difference from the Greenstein method \cite{Greenstein769521} is the use of minimum stationary windows, ensuring that large-scale variations do not affect the small-scale fading statistics.



\vspace{1em}
\subsubsection{\textbf{RMS Delay Spread}}
The RMS delay spread $S_\tau$ describes how multipath signals spread in time within each stationary window $L_{\min}$. For every window centered at time index $t_i$, the time-domain Channel Impulse Response (CIR) $h_m(t_i,\tau)$ is obtained by applying an Inverse Fast Fourier transform (IFFT) across the subcarrier dimension of the measured channel $\mathbf{H}[n]$. 
The corresponding PDP is then calculated by combining the CIR magnitudes over all receive antennas and samples within the local WSS window:
\begin{equation}
P_h(t_i,\tau) = \frac{1}{M W}\sum_{k=i}^{i+W-1}\sum_{m=1}^{M}
\big|h_m(t_k,\tau)\big|^2,
\label{eq:avg_pdp_rms}
\end{equation}
where $W$ denotes the number of samples corresponding to the minimum stationarity distance $L_{\min}$. The RMS delay spread $S_\tau(t_i)$ is subsequently obtained from the locally averaged PDP as~\cite{choi2010generation}:
\begin{equation}
S_\tau(t_i) = 
\sqrt{
\frac{\sum_{\tau=0}^{N_\tau} P_h(t_i,\tau)\,\tau^2}{P_m(t_i)} 
- 
T_m^2(t_i)},
\label{eq:rms_delay_final}
\end{equation}
where
\begin{align}
P_m(t_i) &= \sum_{\tau=0}^{N_\tau} P_h(t_i,\tau), 
\label{eq:Pmean_final}\\[3pt]
T_m(t_i) &= \frac{\sum_{\tau=0}^{N_\tau} P_h(t_i,\tau)\,\tau}{P_m(t_i)}.
\label{eq:Tmean_final}
\end{align}

\vspace{1em}
\subsubsection{\textbf{Spectral Efficiency}}
Assuming Maximum Ratio Combining (MRC) across $M$ receive antennas and a per-subcarrier signal-to-noise ratio (SNR) denoted by $\xi$, the local SE within each stationary window is given by
\begin{equation}
\eta(t_i) = \frac{1}{F}\sum_{f=1}^{F} 
\log_2\!\Big(1 + \xi \cdot \sum_{m=1}^{M} |H_m(t_i,f)|^2\Big),
\label{eq:SE_MRC}
\end{equation}
where the inner summation represents the post-combining power obtained through MRC for subcarrier $f$.


\section{Experimental setup}
\label{sec3}

This section describes the experimental setup. We first describe the hardware, and then we introduce the experimental environment and trajectories. 

\begin{figure}[t!]
    \centering
    \includegraphics[width=0.8\linewidth]{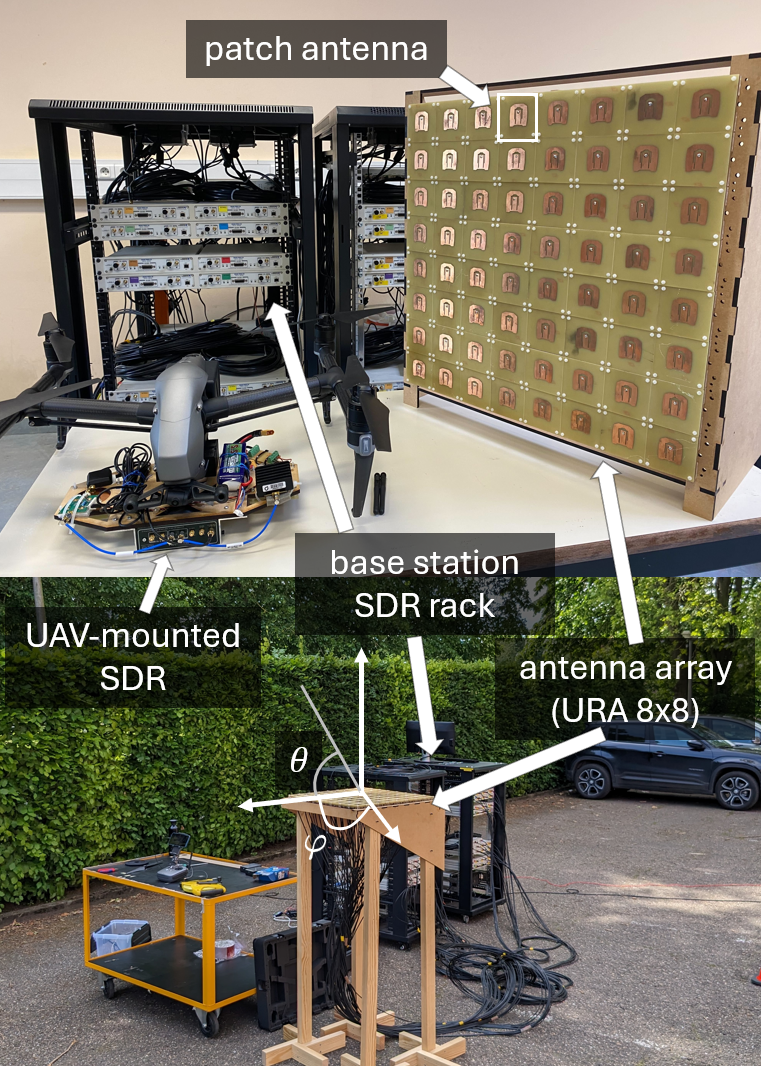}
    \caption{Overview of the measurement equipment and its deployment in the field. The GBS antenna array with upward bore perpendicular to the ground plane is located in a parking lot. The elevation and azimuth angles, $\theta$ and $\varphi$.}
    \label{fig:equipment}
\end{figure}

\subsection{Measurement equipment}
A picture of the used measurement equipment is shown in Fig. \ref{fig:equipment}, where a GBS is equipped with 64 patch antennas specifically designed for an antenna array setup \cite{Chen2017finite}. 
Each antenna has a half power beam width of 75 degrees by 75 degrees.
The antennas are arranged in a Uniform Rectangular Array (URA) of eight-by-eight antennas. The antenna array is positioned 1.2~m above the ground, with its boresight pointed upwards towards the zenith, under a perpendicular angle with the ground, to mimic a car-mounted configuration.

The 64 antennas connect to 32 Software Defined Radios (SDRs). 
These SDRs are kept in sync by a shared 10~MHz input reference clock generated by a GPS Disciplined Oscillator (GPSDO). The center frequency used is 2.61~GHz, and the bandwidth is 18~MHz. The GBS in the setup uses an LTE-based Time Division Duplexing (TDD) frame structure with an OFDM signal. All the SDRs collect IQ samples, and a central system aggregates these IQ samples and performs channel estimation every 1~ms. In the end, the GBS writes these channel estimations to a database file. The Key Parameters of the measurement campaign are provided in Table \ref{measurement_parameters}.

The mobile terminal consists of an E320 SDR mounted on a DJI Inspire 2 UAV or AUE. A single patch antenna is installed underneath the AUE, oriented downward toward the ground to ensure maximum coupling with the ground-based receiver and to minimize interference from multipath components originating from above the UAV. The mobile station transmits an OFDM pilot symbol every 1~ms at a power level of 30~dBm, amplified by an external power amplifier with a gain of 30~dB. The E320 uses a built-in GPSDO to synchronize its internal clock to the same clock as the GBS. The collected channel estimations are spatially localized using the drone GPS coordinates in post-processing.

\begin{figure*}[!t]
    \centering

    \subfloat[Environment with aerial view of horizontal and vertical UAV trajectories.]{%
        \includegraphics[width=0.41\textwidth]{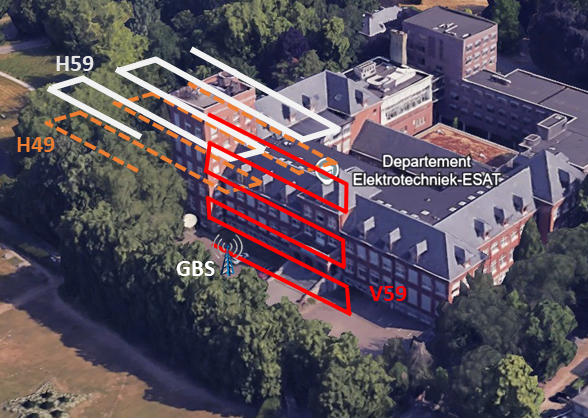}
        \label{fig2a}
    }
    \hfill
    \subfloat[3D view of UAV flight trajectories. Black: H59, Orange: H49, Blue: V59.]{%
        \includegraphics[width=0.45\textwidth]{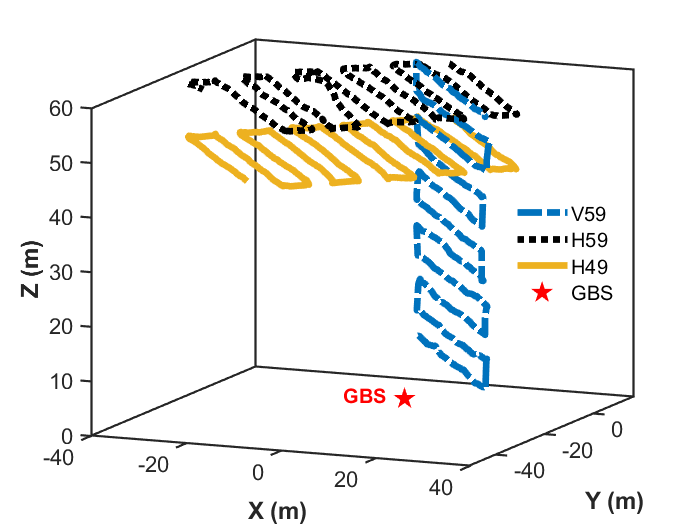}
        \label{fig2b}
    }

    \caption{Illustration of UAV trajectories during the measurement campaign. Aerial view showing H49, H59, and V59 trajectories. (b) 3D MATLAB rendering of the same trajectories relative to the ground GBS (red star).}
    \label{Traj}
\end{figure*}

\subsection{Environment and trajectories}

The GBS is located in a parking lot, 17 m away from the 25 m-tall building and 12 m from the 25 m-high tree line. The parking lot environment during the measurements, conducted on a sunny, clear day in late spring, is shown in Fig. \ref{fig:equipment}. The information for each trajectory is detailed in Table \ref{tab:trajectory_info} and visualized in Fig.~\ref{fig2a}. The UAV is computer-controlled to follow predefined GPS waypoint paths, maintaining a fixed heading aligned with the base station’s polarization. Each trajectory follows a block wave scanning pattern in either a horizontal (H) or vertical (V) plane. Due to significant vegetation-induced attenuation, no measurements could be obtained behind the tree line, as visualized in Fig. \ref{fig2a}. Therefore, the measured propagation conditions are mainly in LoS. 

\begin{table}[!t]
\centering
\caption{Key Parameters of the Measurement Campaign}
\label{measurement_parameters}
\renewcommand{\arraystretch}{1.15}
\begin{tabular}{l c}
\hline
\textbf{Parameter} & \textbf{Value / Description} \\
\hline
Carrier frequency & 2.61 GHz \\

Total bandwidth & 18 MHz \\

Number of subcarriers & 100 \\

Transmit antennas (GBS) & 64-element URA \\

Receive antennas (UAV) & 1 \\

AUE altitude range & 0 to 59 m \\

Number of trajectories & 3 \\

Environment & Suburban (mild scattering) \\

CSI capture rate & 1 kHz\\
\hline
\end{tabular}
\end{table}

In total, we consider three different UAV trajectories, given in Table~\ref{tab:trajectory_info}, to examine various spatial and angular propagation characteristics. The UAV operated at heights up to 59 m, approximately twice the surrounding building height (25~m), to maintain LoS while capturing altitude-dependent multipath variations and corresponding changes in the K-factor.

\begin{itemize}
    \item \textbf{Trajectory 1 (H49):} A horizontal zig-zag pattern at a constant altitude of approximately 49~m, designed to isolate azimuthal and horizontal distance effects while restricting the variation in elevation angles.
    
    \item \textbf{Trajectory 2 (H59):} Similar horizontal zig-zag motion at a higher altitude of 59 m.
    
    \item \textbf{Trajectory 3 (V59):} A vertical zig-zag ascent from ground level to approximately 59~m, intended to analyze combined variations in elevation angle, azimuth, three-dimensional distance, and multipath characteristics.
\end{itemize}

The combined design enables analysis of both horizontal and vertical mobility patterns. Fig.~\ref{fig2b} visualizes the 3D UAV trajectories in the MATLAB environment, with the GBS marked by a red star. The zig-zag paths span diverse positions, angles, and distances, enabling robust multi-domain channel characterization.

\begin{table}[t!]
    \centering
    \renewcommand{\arraystretch}{1.15}
    \caption{Trajectory information}
    \label{tab:trajectory_info}
   \begin{tabular}{c*{4}{c}}
\hline
\multirow{2}{*}{Traj.} & \multirow{2}{*}{Samples} & \multirow{2}{*}{Height (m)} & {Mean} & {Scan} \\ 
& & & {velocity (m/s)} & {pattern}\\
\hline
1 & 149970 & 49\textsuperscript{(a)} & 4.06 & H\\ 
2 & 149950 & 59\textsuperscript{(a)} & 3.84 & H\\ 
3 & 149945  & 10--59\textsuperscript{(b)} & 2.38 & V\\ 
\hline
\end{tabular}
\begin{flushright}
(a) fixed height, (b) vertical scan pattern 
\end{flushright}
\end{table}

\begin{figure*}[t]
\centering

\subfloat[H49: Power vs Elevation]{%
  \includegraphics[width=0.3\textwidth,trim=2 2 2 2,clip]{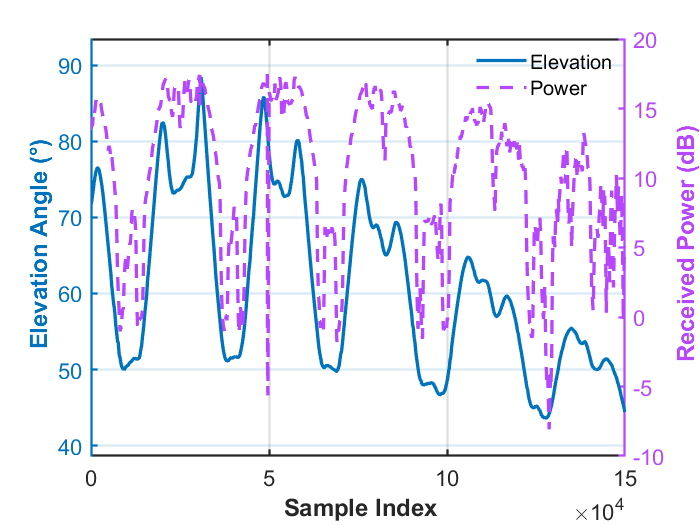}}
\hspace{0.015\textwidth}
\subfloat[H49: Power vs Azimuth]{%
  \includegraphics[width=0.3\textwidth,trim=2 2 2 2,clip]{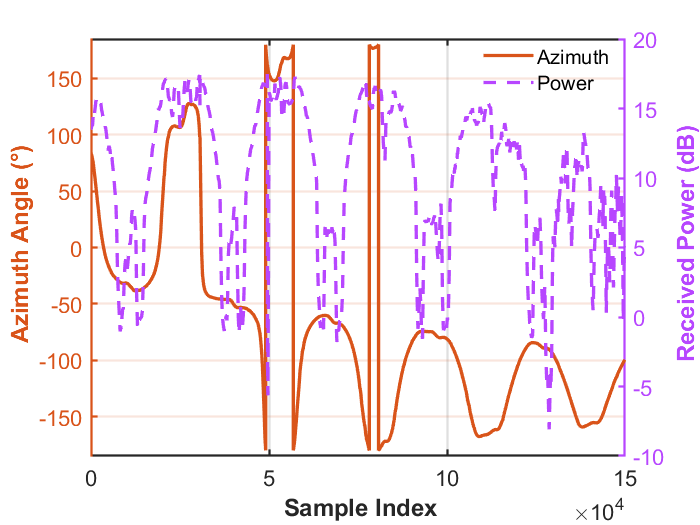}}
\hspace{0.015\textwidth}
\subfloat[H49: Power vs Distance]{%
  \includegraphics[width=0.3\textwidth,trim=2 2 2 2,clip]{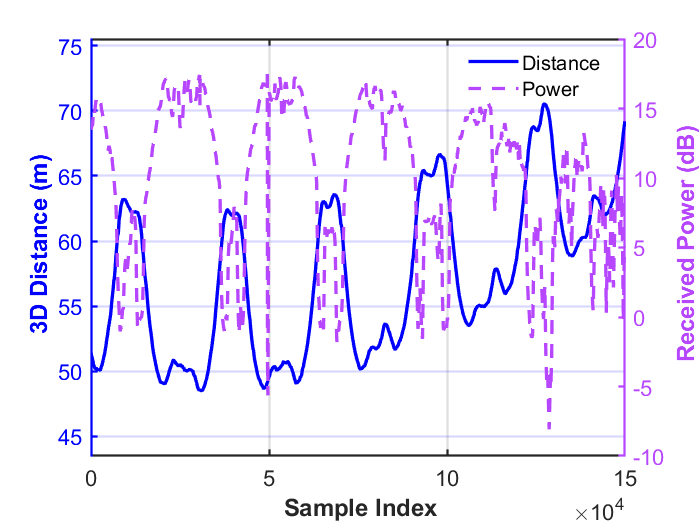}}

\vspace{4pt}

\subfloat[H59: Power vs Elevation]{%
  \includegraphics[width=0.3\textwidth,trim=2 2 2 2,clip]{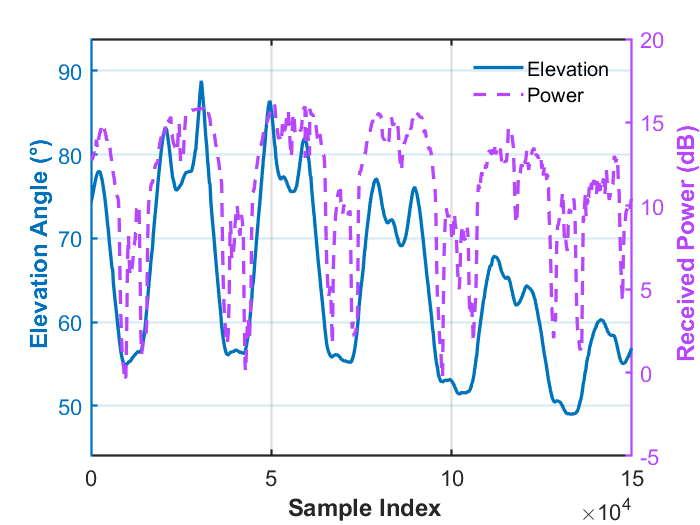}}
\hspace{0.015\textwidth}
\subfloat[H59: Power vs Azimuth]{%
  \includegraphics[width=0.3\textwidth,trim=2 2 2 2,clip]{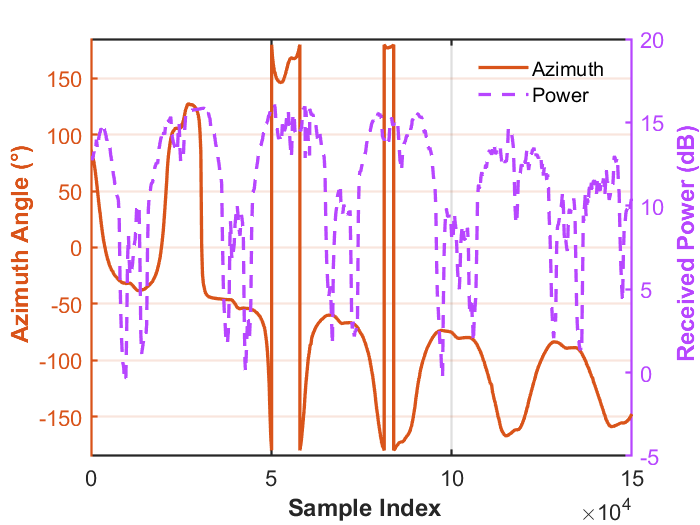}}
\hspace{0.015\textwidth}
\subfloat[H59: Power vs Distance]{%
  \includegraphics[width=0.3\textwidth,trim=2 2 2 2,clip]{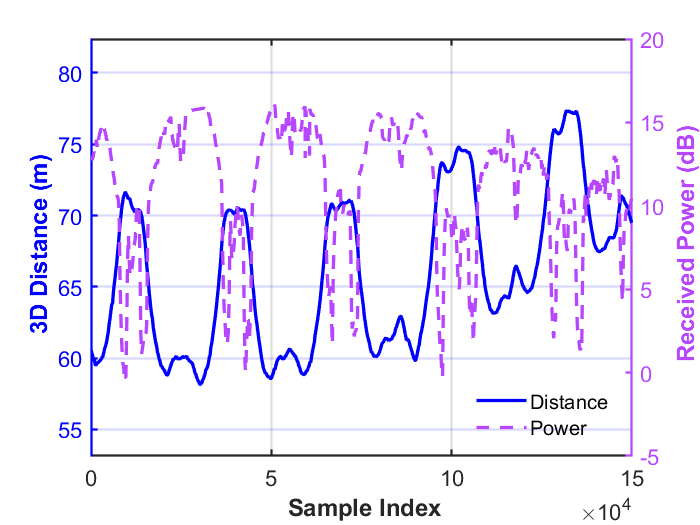}}

\vspace{4pt}

\subfloat[V59: Power vs Elevation]{%
  \includegraphics[width=0.3\textwidth,trim=2 2 2 2,clip]{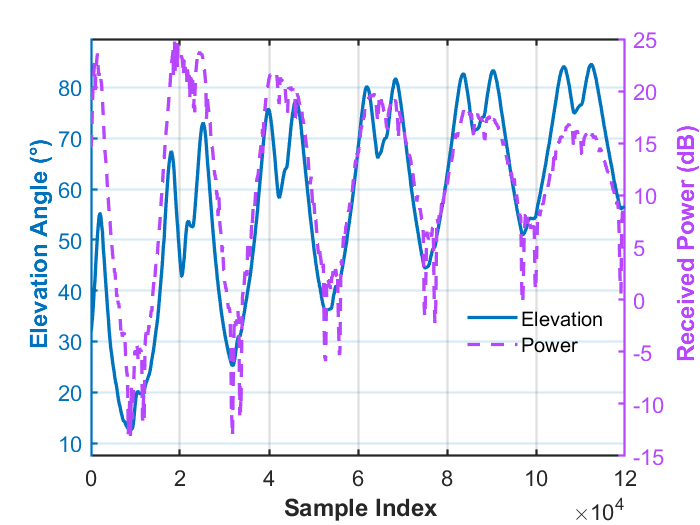}}
\hspace{0.015\textwidth}
\subfloat[V59: Power vs Azimuth\label{3gFig}]{%
  \includegraphics[width=0.3\textwidth,trim=2 2 2 2,clip]{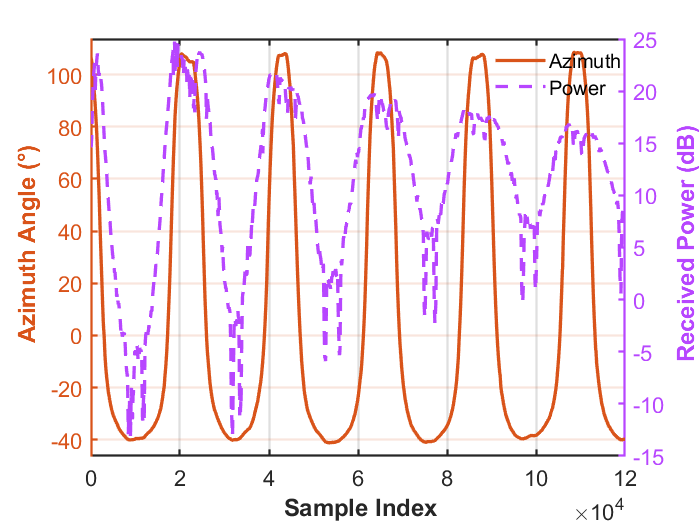}}
\hspace{0.015\textwidth}
\subfloat[V59: Power vs Distance]{%
  \includegraphics[width=0.3\textwidth,trim=2 2 2 2,clip]{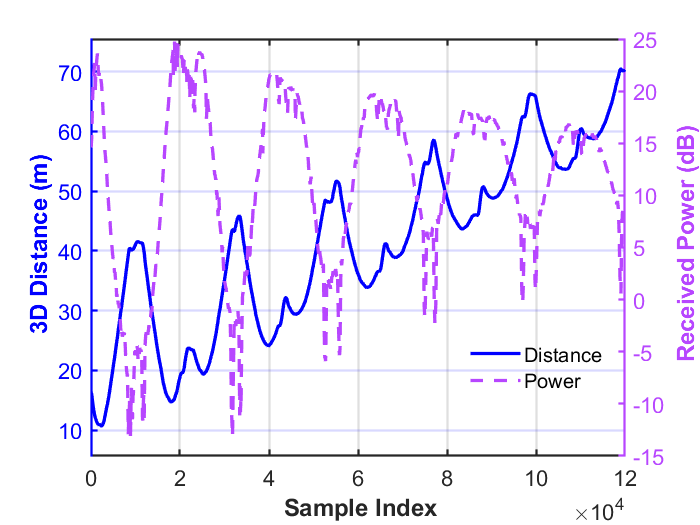}}

\caption{Correlation of received power with elevation angle, azimuth angle, and 3D distance for different UAV trajectories (H49, H59, V59).}
\label{Corrfig}
\end{figure*}

\begin{figure}[!t]
    \centering
    \includegraphics[width=.7\linewidth]{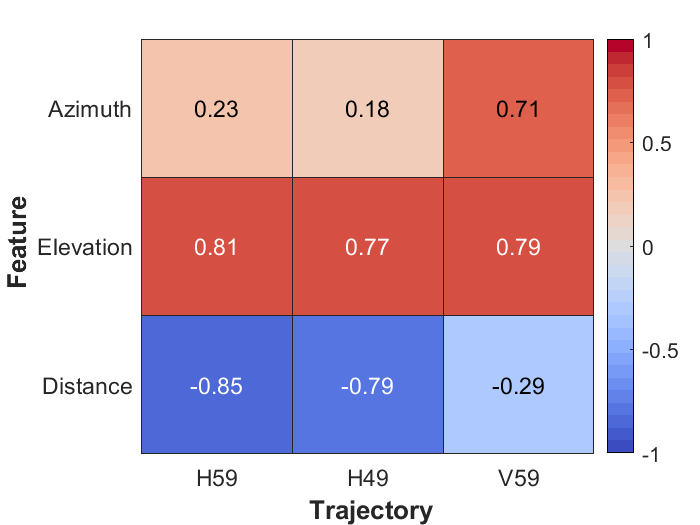}
    \caption{Correlation of received power with elevation, azimuth, and 3D distance across all trajectories (H59, H49, V59). }
    \label{Combined_Corr}
\end{figure}

\section{Geometry-Driven Channel Behavior}
\label{sec4}

In this paper, we analyze three specific flight trajectories to gain a thorough understanding of the propagation dynamics of A2G channels in a suburban environment, as presented in Section~\ref{sec3}. Two horizontal zig-zag flights were performed at fixed altitudes of 49~m (H49) and 59~m (H59), while one vertical ascent trajectory (V59) gradually increased altitude up to 59~m with horizontal scanning patterns. The zig-zag design was selected to mimic practical AUE operations such as surveillance or search and rescue, while simultaneously inducing spatial and angular diversity in both azimuth and elevation domains. Together, these trajectories cover a diverse range of distances and angles, yielding a comprehensive dataset for characterizing multi-domain channels.


\subsection{Large-Scale Power Variations}
Fig. \ref{Corrfig} illustrates the dependence of received power on the UAV’s geometric parameters, including distance, azimuth, and elevation angles for the three trajectories. For horizontal flights (H49/H59), which are mainly in LoS, we observe that power closely follows distance (anti-correlation) and elevation variations, showing smooth trends with some fluctuations due to multipath. The azimuth power plots show periodic changes that match the zig-zag flight path. These fluctuations mainly arise from the antenna’s gain pattern and weak multipath effects.


For the vertical flight (V59), the relationship between power and geometry is different. In this case, the received power follows the azimuth variation more closely than the distance. The zig-zag flight path creates repeated azimuthal shifts. These shifts result in power fluctuations due to antenna gain patterns and multipath effects, which are stronger at lower and mid-altitudes before becoming more stable at higher elevation angles, as shown in Fig. \ref{3gFig}. The strong azimuth-power link demonstrates how UAV yaw and lateral movement impact signal quality, highlighting the importance of angle-aware beam management in AUE operations.

\begin{figure*}[t]
\centering
\subfloat[H49 trajectory \label{fig5a}]{%
  \includegraphics[width=0.3\textwidth,trim=2 2 2 2,clip]{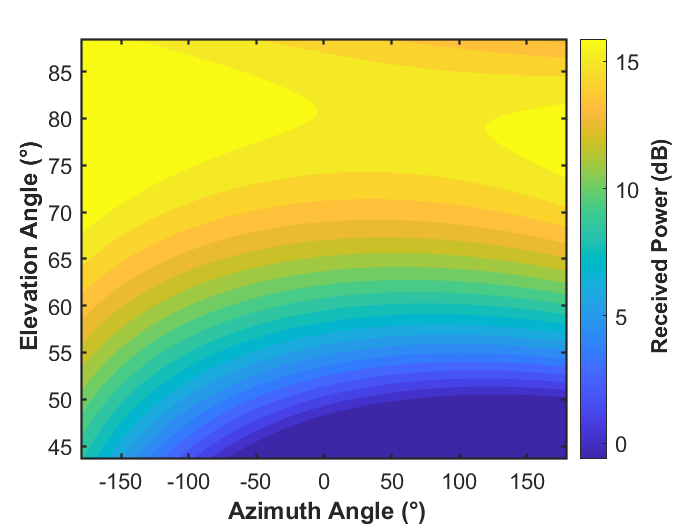}}
\hspace{0.015\textwidth}
\subfloat[H59 trajectory \label{fig5b}]{%
  \includegraphics[width=0.3\textwidth,trim=2 2 2 2,clip]{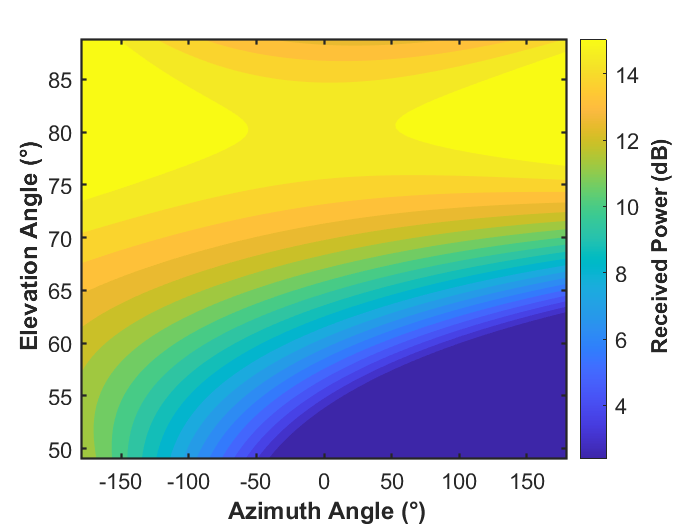}}
\hspace{0.015\textwidth}
\subfloat[V59 trajectory \label{fig5c}]{%
  \includegraphics[width=0.3\textwidth,trim=2 2 2 2,clip]{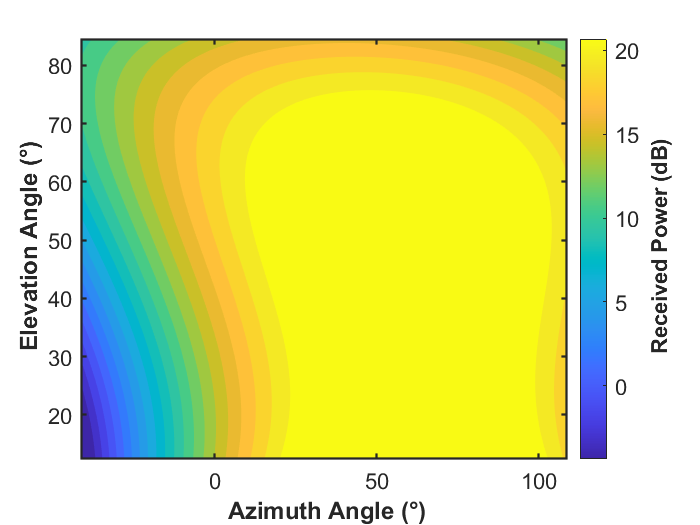}}

\vspace{4pt}
\caption{Measured received power as a function of azimuth and elevation angles for three UAV trajectories. These heatmaps highlight angular power distributions with variations introduced by the mobility and geometry of UAVs.}
\label{3Dpower}
\end{figure*}

Fig. \ref{Combined_Corr} summarizes the correlation between received power and geometric features across all trajectories, showing that elevation has the strongest correlation (0.81, 0.77, 0.79 for H59, H49, and V59), making it the strongest predictor of received power. In contrast, distance remains strongly anticorrelated in the horizontal flights, but only a weak one (-0.29) in the vertical case. However, the correlation between distance and power is minimal. Azimuth becomes more relevant in vertical trajectories ($0.71$), especially at lower heights. Overall, elevation provides the most stable indicator of received power, while distance and azimuth dependencies vary depending on the UAV’s flight pattern. Fig. \ref{3Dpower} supports this observation: in the horizontal flights, power variations are mainly driven by elevation, whereas in the vertical flight, azimuth has a stronger influence. 

\begin{figure*}[t]
\centering
\subfloat[H49: Fading envelope CDF with fits]{%
  \includegraphics[width=0.3\textwidth,trim=2 2 2 2,clip]{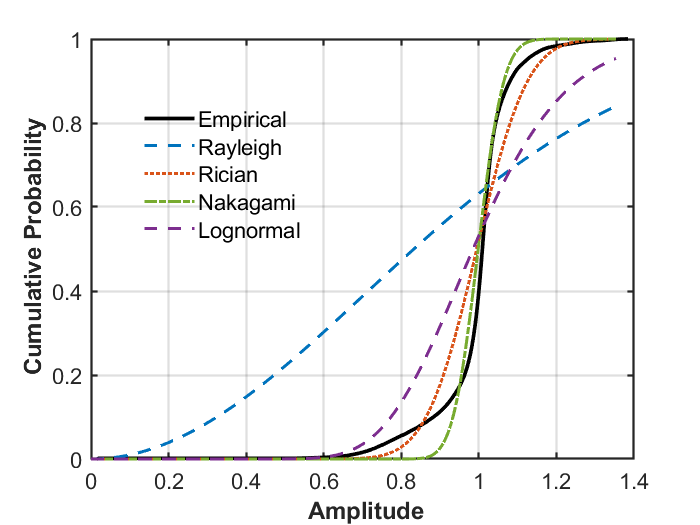}}
\hspace{0.015\textwidth}
\subfloat[H59: Fading envelope CDF with fits]{%
  \includegraphics[width=0.3\textwidth,trim=2 2 2 2,clip]{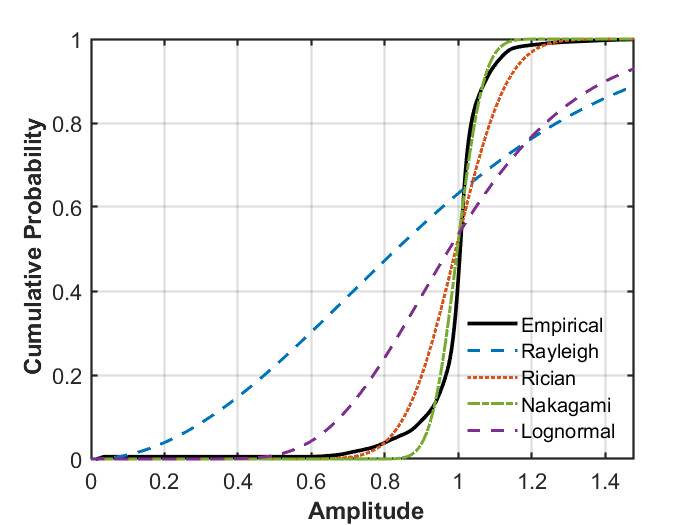}}
\hspace{0.015\textwidth}
\subfloat[V59: Fading envelope CDF with fits]{%
  \includegraphics[width=0.3\textwidth,trim=2 2 2 2,clip]{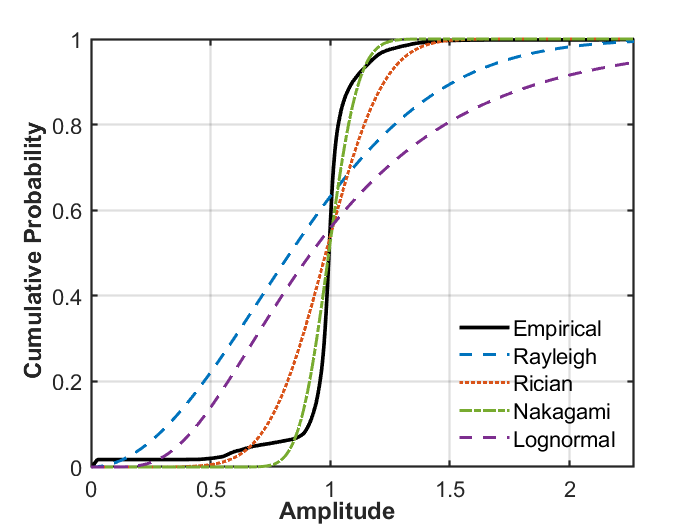}}
\caption{Empirical CDFs of normalized small-scale fading envelopes for all three UAV trajectories, compared against Rayleigh, Rician, Nakagami, and Lognormal distributions. }
\label{SSF_CDFs}
\end{figure*}

\subsection{Small-Scale Fading}

To characterize the short-term fading behavior of the A2G channel, we analyze the distribution of the small-scale fading envelope $a_{\mathrm{SSF}}[n]$, extracted using the $60\lambda$ moving average method described in Section~\ref{sec2}. The primary objective of using this window length is to effectively eliminate large-scale variations resulting from path loss and shadowing, while preserving rapid amplitude fluctuations associated with small-scale multipath fading. 

We fit the Cumulative Distribution Functions (CDFs) of the resulting fading envelope in Fig.~\ref{SSF_CDFs} against the four widely used fading models (Rayleigh, Rician, Nakagami, and Lognormal) to evaluate the model fitness using the Kolmogorov–Smirnov (KS) distance~\cite{massey1951ks}, which quantifies the maximum difference between the empirical and theoretical CDFs. Table~\ref{KSdist} plots the KS distances, where smaller KS values indicate better agreement between the measured and modeled data.

\begin{table}[!t]
\centering
\caption{KS distances for fading models across all UAV trajectories.}
\label{KSdist}
\renewcommand{\arraystretch}{1.15}
\begin{tabular}{lcccc}
\hline
\textbf{Trajectory} & \textbf{Rayleigh} & \textbf{Rician} & \textbf{Nakagami} & \textbf{Lognormal} \\
\hline
H49 & 0.443 & 0.187 & \cellcolor{maxshade}\textbf{0.122} & 0.253 \\
H59 & 0.465 & 0.205 & \cellcolor{maxshade}\textbf{0.127} & 0.311 \\
V59 & 0.468 & 0.267 & \cellcolor{maxshade}\textbf{0.152} & 0.397 \\
\hline
\end{tabular}
\end{table}

Unlike the findings in \cite{gaertner2007characterizing, cai2021characterizing}, where the Rician distribution provided the best fit followed by Nakagami, our results show that the Nakagami distribution provides the lowest KS distance in all trajectories, followed by the Rician distribution. This contrast can be explained by the differing propagation environments and flight geometries. The cited studies focused on dense urban or mixed cluttered environments with stronger multipath and higher scattering, which favor Rician behavior. In contrast, our suburban scenario features a dominant LoS component combined with numerous weak multipath reflections from buildings, trees, and the ground. The main advantage of the Nakagami model is its flexibility due to its shape parameter $m$, which allows it to represent conditions ranging from Rayleigh (no LoS) to Rician (dominant LoS), making it well-suited for partially LoS environments. The second best fit of the Rician model further supports the presence of a stable LoS path with mild diffuse scattering.

In contrast, the Rayleigh model, assuming fully non-LoS conditions, underestimates received power and deviates from the empirical CDF, while the Lognormal model fits only the distribution tail. Relatively high KS distances ($>$0.12) arise from real-world imperfections, as the UAV channel is non-stationary with geometry-dependent LoS and scattered components. Continuous elevation and azimuth changes, along with noise and antenna effects, further distort the fading envelope, leading to slight deviations from ideal theoretical models, as observed in outdoor flight measurements. 

While the overall fading follows a Nakagami distribution, this represents a mix of different channel conditions along the trajectory. Within each locally stationary region, the fading behaves approximately Rician with a varying K-factor that captures the relative strength of the LoS component.



\begin{figure}[!t]
    \centering
    \includegraphics[width=.75\linewidth,trim=2 2 2 2,clip]{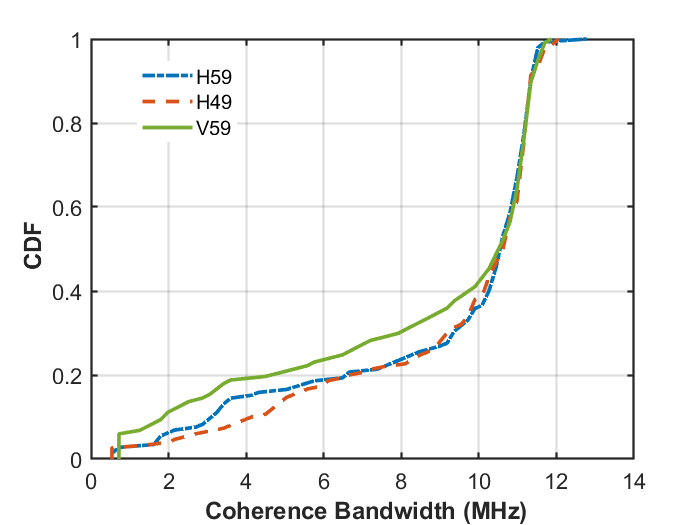}
    \caption{Empirical CDFs of $B_{\mathrm{coh}}$ for the three UAV trajectories. }
    \label{Bcoh}
    \vspace{-.5em}
\end{figure}

\section{Stationarity and Performance Evaluation}
\label{statSec}

\subsection{Frequency Stationarity}
The frequency-domain stationarity of the UAV channels was evaluated using the coherence bandwidth $B_{\mathrm{coh}}$, calculated using \eqref{eq:coh_bw}. Fig. \ref{Bcoh} shows the empirical CDFs of coherence bandwidth for all trajectories. Across the measurements, most of the $B_{\mathrm{coh}}$ values mainly fall between 2 and 12 MHz, with median values around 7.2$\to$7.6 MHz for all the flights (H49, H59) and slightly lower for the vertical flight (V59). The left-shifted and narrower distribution in V59 indicates stronger frequency selectivity, mainly due to greater elevation and distance changes during ascent. Overall, the results indicate that the measured A2G channels maintain relatively wide $B_{\mathrm{coh}}$ due to dominant LoS propagation and limited multipath in the suburban environment.

To verify the stability of channel statistics across frequency, we divide the 18 MHz band into five subbands, each sufficiently narrow to lie within the typical coherence bandwidth range. The evaluation shows that the first-order statistics remained nearly constant, with average Root Mean Square Error (RMSE) values of 7.4\% for the mean and 1.1\% for the standard deviation across the subbands. Therefore, the full 18 MHz bandwidth is used for the temporal stationarity analysis.

\begin{table}[!t]
\caption{Stationarity summary by trajectory. 
Highlighted cells denote the minimum (blue shading) and maximum (red shading) across all trajectories for each metric.}
\label{summaryTab}
\centering
\renewcommand{\arraystretch}{1.15}
\begin{tabular}{l l c c c c}
\hline
\multirow{2}{*}{Traj.} & \multirow{2}{*}{Geometric Features} 
& \multicolumn{4}{c}{Normalized Stationarity Span} \\
\cline{3-6}
 &  & Min & Max & Mean & STD \\
\hline
\multirow{3}{*}{H49} 
 & Elevation  & 0.0045 & 0.2534 & 0.1203 & 0.0575 \\
 & Azimuth    & 0.0023 & 0.4318 & 0.0499 & 0.0510 \\
 & Distance   & 0.0093 & 0.0323 & 0.0173 & \cellcolor{minshade}\textbf{0.0036} \\
\hline
\multirow{3}{*}{H59} 
 & Elevation  & \cellcolor{minshade}\textbf{0.0018} & \cellcolor{maxshade}\textbf{0.2760} & \cellcolor{maxshade}\textbf{0.1280} & 0.0613 \\
 & Azimuth    & \cellcolor{minshade}\textbf{0.0009} & 0.3470 & 0.0510 & \cellcolor{minshade}\textbf{0.0462} \\
 & Distance   & \cellcolor{minshade}\textbf{0.0005} & 0.0375 & \cellcolor{maxshade}\textbf{0.0196} & 0.0052 \\
\hline
\multirow{3}{*}{V59} 
 & Elevation  & 0.0040 & 0.2062 & 0.1156 & \cellcolor{minshade}\textbf{0.0416} \\
 & Azimuth    & 0.0024 & \cellcolor{maxshade}\textbf{0.5077} & \cellcolor{maxshade}\textbf{0.1092} & 0.1141 \\
 & Distance   & 0.0013 & \cellcolor{maxshade}\textbf{0.0420} & 0.0165 & 0.0074 \\
\hline
\end{tabular}
\end{table}

\subsection{Temporal and Angular Stationarity}

\begin{figure*}[t]
\centering
\subfloat[H49 trajectory]{%
  \includegraphics[width=0.3\textwidth,trim=2 2 2 2,clip]{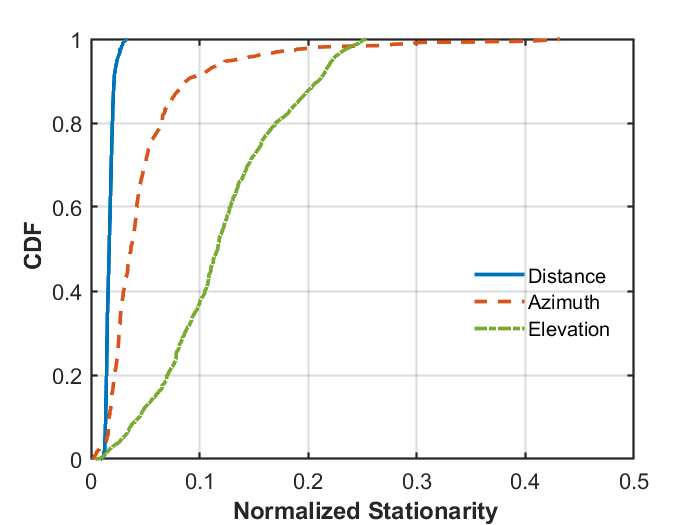}}
\hspace{0.015\textwidth}
\subfloat[H59 trajectory]{%
  \includegraphics[width=0.3\textwidth,trim=2 2 2 2,clip]{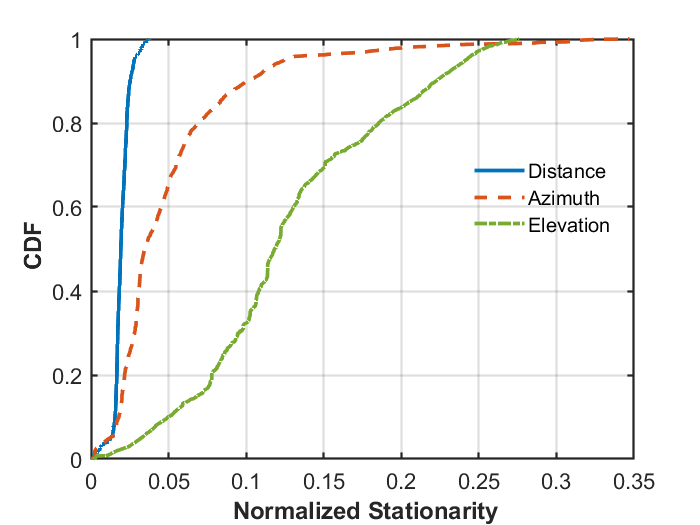}}
\hspace{0.015\textwidth}
\subfloat[V59 trajectory]{%
  \includegraphics[width=0.3\textwidth,trim=2 2 2 2,clip]{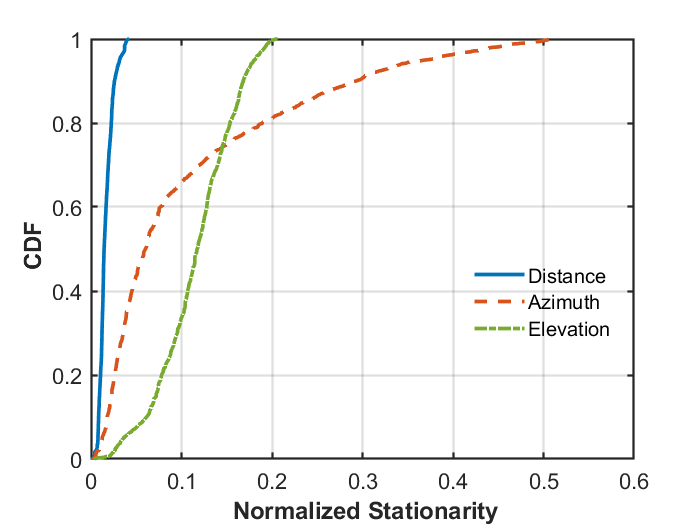}}

\vspace{2pt}
\caption{Empirical CDFs of normalized stationarity regions for azimuth, elevation, and 3D distance along three UAV trajectories.}
\label{CDF_St_1D}
\end{figure*}

Fig. \ref{CDF_St_1D} plots the empirical CDFs of normalized stationarity regions for azimuth, elevation, and 3D distance along three UAV trajectories. Here, the CMD-based stationarity spans $D_{\text{stat}}(t_i)$, $A_{\text{az}}(t_i)$, and $A_{\text{el}}(t_i)$ are normalized by their respective total range. The results show that in both horizontal flights, azimuth exhibits the shortest and most variable stationarity spans (mean values of 0.0499 and 0.0510), confirming that the UAV’s side-to-side zig-zag movement induces rapid angular decorrelation. This fast variation in azimuth causes the spatial correlation matrix to change quickly, especially when the UAV periodically turns at the ends of each leg. Elevation shows longer stationarity spans (mean 0.1203 and 0.1280), as the UAV altitude remains nearly constant during horizontal flight, leading to more stable vertical channel statistics. In contrast, distance shows the smallest spans (mean 0.0173 and 0.0196) and very low Standard Deviation (STD), indicating that even small changes in range during horizontal movement quickly affect the correlation structure due to the LoS phase evolution along the trajectory. These findings suggest that angular changes, particularly in azimuth, have a more significant impact on non-stationarity during horizontal flight, while distance exhibits the shortest but most consistent stationarity behavior. Table \ref{summaryTab} further summarizes the minimum, maximum, mean, and standard deviation of these normalized stationarity spans across all trajectories.

In the vertical trajectory (V59), a different trend is observed. The azimuth shows the largest and most variable stationarity spans (mean 0.1092, max 0.5077), reflecting that the UAV’s primarily upward motion causes minimal change in the horizontal direction. Elevation (mean 0.1156) and distance (mean 0.0165) exhibit relatively shorter spans, as continuous altitude changes directly affect both the propagation range and elevation angle. In essence, rapid altitude changes cause the elevation and distance domains to decorrelate quickly. In contrast, the azimuth remains more stationary because the UAV’s horizontal direction changes less compared to elevation and distance.


\subsection{Rician K-Factor}

To examine the strength of the LoS component in the A2G channel, we estimate the Rician K-factor from the small-scale fading envelope using a moment-based method~\cite{Greenstein769521}. However, we compute the K-factor values within each quasi-stationary window defined by the minimum stationarity distance $L_{\min}$ to minimize the influence of large-scale variations. Although the overall fading follows a Nakagami distribution, the K-factor remains a valuable metric for describing LoS dominance. Each quasi-stationary window exhibits approximately Rician behavior, with a unique K value that changes as the UAV's geometry and environment evolve. Tracking this variation offers insight into how the LoS and diffuse components vary with angle and altitude.

\begin{figure*}[t]
\centering
\subfloat[H49 trajectory \label{7a}]{%
  \includegraphics[width=0.3\textwidth,trim=2 2 2 2,clip]{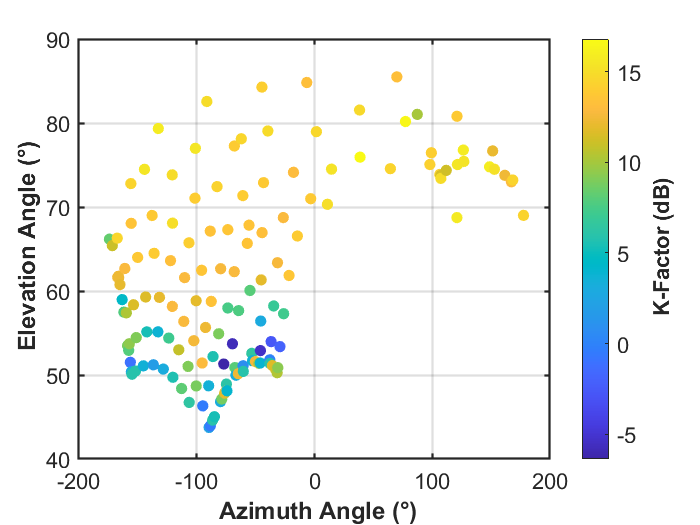}}
\hspace{0.015\textwidth}
\subfloat[H59 trajectory \label{7b}]{%
  \includegraphics[width=0.3\textwidth,trim=2 2 2 2,clip]{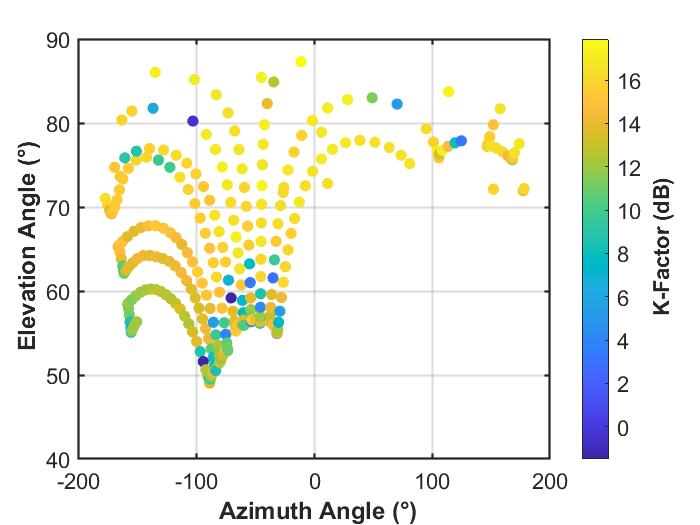}}
\hspace{0.015\textwidth}
\subfloat[V59 trajectory]{%
  \includegraphics[width=0.3\textwidth,trim=2 2 2 2,clip]{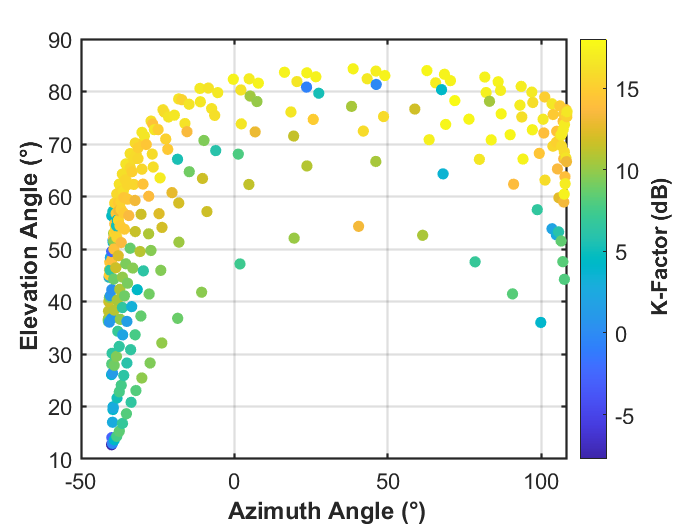}}

\caption{Estimated K-factor distribution as a function of azimuth and elevation angles for (a) H49, (b) H59, and (c) V59 trajectories. Each point represents a local K-factor estimate, with color indicating magnitude. }
\label{kfac}
\end{figure*}

Fig.~\ref{kfac} shows the variation of K-factor as a function of geometric angles ($\theta, \varphi$), where yellow color indicates higher values. Overall, we observe that the K-factor exhibits a positive trend with elevation angle, which is associated with the AUE altitude. We can observe this trend in the vertical trajectory (V59), where higher K-factor values of $\approx$ 15~dB are observed with higher elevation angles when the AUE ascends from ground level to approximately 59~m, which is consistent with findings in \cite{qiu2017low}. The primary reason behind this is that higher elevation angles reduce the chance of blockage from trees or buildings, providing a clearer LoS path and fewer strong multipath components, leading to a higher K-factor. In contrast, lower altitudes experience more significant fluctuations in the K-factor due to intermittent shadowing and angular variations introduced by zig-zag UAV movement.

\begin{figure}[!t]
    \centering
    \includegraphics[width=1\linewidth]{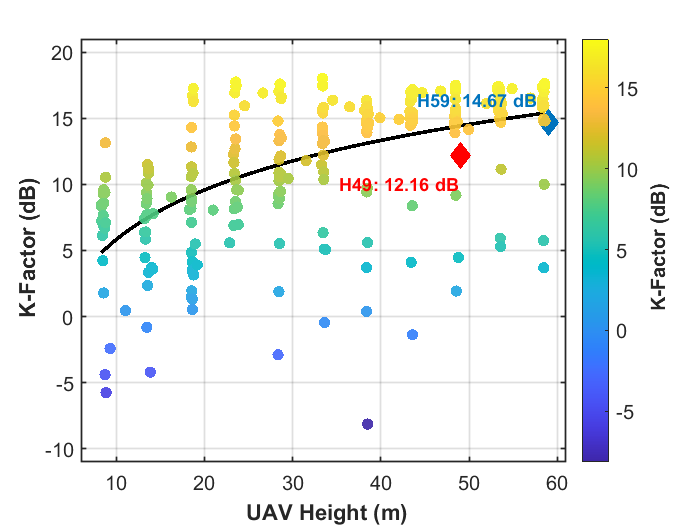}
    \caption{Variation of the K-factor with UAV altitude in V59 trajectory. Each point represents a local estimate of the K-factor, with a Log fit (black) showing a slight upward trend. Markers highlight average values for the H49 and H59 horizontal flight trajectories.}
    \label{Kfac_fit}
\end{figure}

The influence of height on the K-factor is more evident in Figs \ref{7a} and \ref{7b}, where both the horizontal trajectories yield similar K-factor trends. However, H59 exhibits higher average K-factor values, suggesting reduced multipath and a dominant LoS path at higher altitudes. Therefore, we model the K-factor as a log-linear function of AUE height in Fig.~\ref{Kfac_fit} to quantify the observed trend with altitude.   The data is fitted using the model given in equation \eqref{logFit}.

\begin{equation}
\label{logFit}
   K(h) = a \cdot \ln(h) + b,
\end{equation}

\noindent
where $h$ is the AUE height (in meters), while $a = 5.414$ and $b = -6.639$ are the fitting parameters. This logarithmic model captures the rapid increase in K-factor at lower heights and its gradual saturation at higher altitudes, reflecting stronger LoS dominance as the UAV rises above suburban obstructions, such as buildings or trees. The red and blue markers in Fig.~\ref{Kfac_fit} indicate the mean K-factor values for the H49 and H59 trajectories, respectively, which align well with the fitted curve. The formulation is restricted to $h \geq 10$ m, as values below this were not included in the measurements. It is worth noting that H49 and H59 employ different $L_{\min}$, leading to different numbers of K-factor samples and plot densities. This confirms that, despite the overall Nakagami envelope behavior, the local fading within each window retains a Rician structure, with the K-factor evolving with geometry and visibility conditions.

\subsection{Linking Channel Metrics to Spectral Efficiency}
To bridge the statistical channel characterization with system-level performance, we analyze the relationship between two key channel metrics, RMS delay spread and K-factor, with the corresponding SE (assuming a fixed $\xi = 20$ dB). All quantities are computed within the minimum stationarity distance $L_{\min}$ (Section~\ref{sec2}) so that each estimate reflects locally WSS conditions. The K-factor is estimated after normalizing windowed power samples to unit mean, which removes large-scale path-loss and shadowing effects to ensure it reflects only the LoS-to-diffuse power ratio. Similarly, the RMS delay spread is calculated from the power-normalized PDP. For brevity, results from the H49 trajectory are omitted, as they follow the same trends as H59. This comparison provides insight into how multipath, LoS dominance, and link budget jointly influence communication capacity in A2G links.

\begin{figure*}[!t]
\centering
\subfloat[SE vs K-Factor (H59)\label{se_k_h59}]{%
  \includegraphics[width=0.48\textwidth,trim=2 2 2 2,clip]{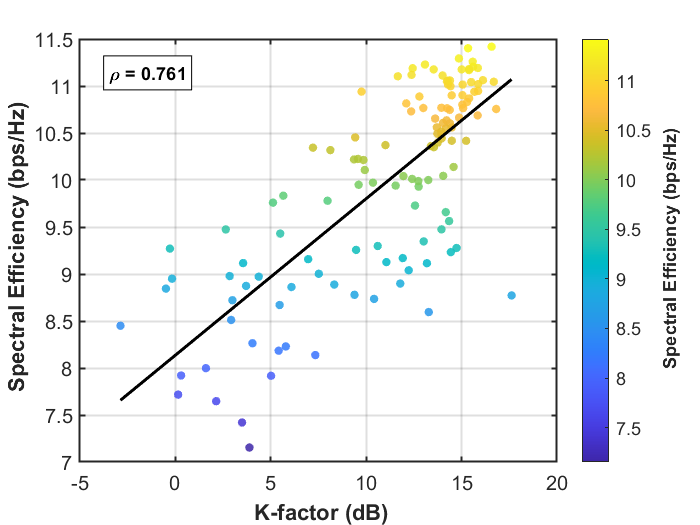}}
\hfill
\hfill
\subfloat[SE vs RMS Delay Spread (H59)\label{se_rms_h59}]{%
  \includegraphics[width=0.48\textwidth,trim=2 2 2 2,clip]{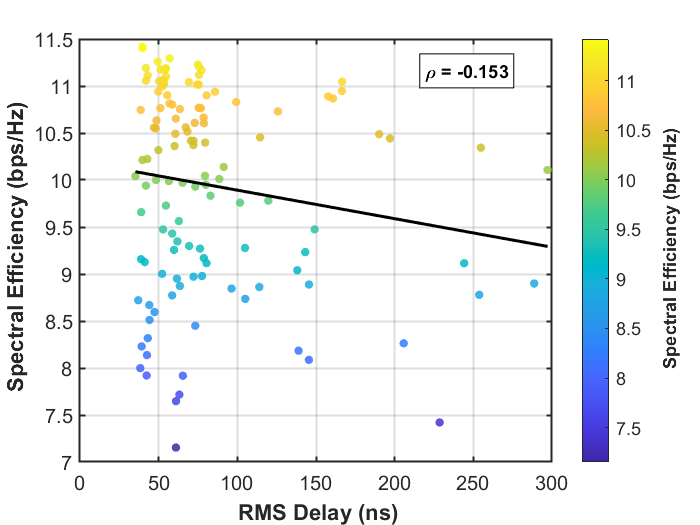}}

\vspace{4pt}

\subfloat[SE vs K-Factor (V59)\label{se_k_v59}]{%
  \includegraphics[width=0.48\textwidth,trim=2 2 2 2,clip]{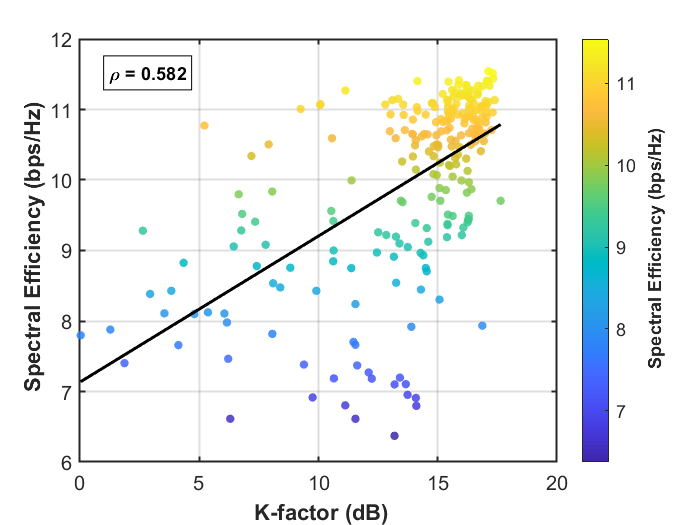}}
\hfill
\hfill
\subfloat[SE vs RMS Delay Spread (V59)\label{se_rms_v59}]{%
  \includegraphics[width=0.48\textwidth,trim=2 2 2 2,clip]{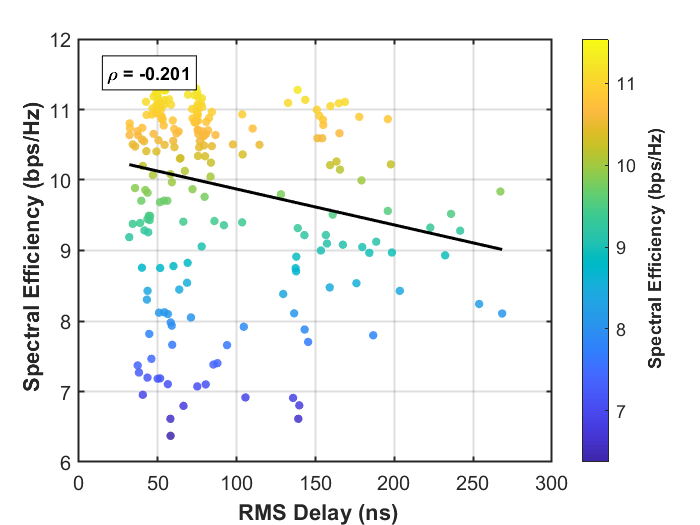}}

\vspace{4pt}
\caption{Spectral efficiency as a function of three key channel metrics for two UAV trajectories (H59 and V59): (a,c) Rician K-factor and (b,d) RMS delay spread. Each subplot shows the correlation coefficient ($\rho$) to quantify the strength and direction of correlation.  }
\label{SE_vs_KPRMS}
\end{figure*}

Figs. \ref{SE_vs_KPRMS}\subref{se_k_h59}–\subref{se_rms_v59} illustrate the trends observed in H59 and V59 trajectories. Across both trajectories, SE shows a moderately strong positive association with the K-factor ($\rho = 0.761$ for H59 and $\rho = 0.582$ for V59), indicating that LoS dominance enhances the channel’s reliability and effective capacity. This effect is more prominent in the horizontal flight, where stable geometry and smoother angular evolution preserve LoS coherence. In contrast, the vertical trajectory shows a larger scatter due to continuous elevation changes and intermittent shadowing, which weakens the instantaneous relationship between the K-factor and SE.

In contrast, the RMS delay spread shows only a weak negative trend with SE: $\rho=-0.153$ for H59 and $\rho=-0.201$ for V59. This weak dependence indicates that delay spread has a limited impact on link capacity in the largely LoS suburban environment. Furthermore, within each stationary window, excess multipath delays remain small, and the 18 MHz OFDM signal efficiently mitigates any residual inter-symbol interference. Overall, once large-scale power variations are removed, delay spread has a minor influence compared to link gain and LoS strength.

These results emphasize that spectral efficiency is primarily driven by LoS strength, while the role of delay dispersion remains secondary under stable suburban LoS conditions. The strong correlation between SE and the K-factor highlights the importance of maintaining high-elevation, low-scattering geometries. In contrast, the weak RMS delay correlation confirms limited time-domain selectivity in such environments. These findings suggest that SE optimization in UAV networks should jointly consider geometry-aware link adaptation, angular stability, and multipath mitigation, rather than relying on a single metric.

\section{Conclusion}
\label{sec5}
In this paper, we present a comprehensive measurement-based study of A2G wireless channel characteristics for low-altitude UAVs operating as AUE. The measurement campaign consists of three different AUE trajectories, where we observe how mobility in different trajectories affects power variation, fading behavior, and stationarity against the geometrical features. Our analysis demonstrates that elevation angle is the most consistent predictor of received power and LoS dominance, with correlation coefficients $\rho_{\text{elev}} \approx 0.77$--$0.81$ across all trajectories and $K_{\text{dB}}$ increasing from about $5$ to over $15$~dB with altitude. At the same time, the Nakagami model is the most suitable for modeling small-scale fading due to suburban shadowing. Additionally, temporal and angular stationarity are shown to be highly sensitive to the UAV’s motion axis, with mean normalized spans of $\approx 0.05$ in azimuth and $\approx 0.12$ in elevation for horizontal flights. We further analyzed SE trends against K-factor and RMS delay spread, revealing their combined impact on link performance. These insights provide critical guidance for designing future NTN-aware UAV communication protocols and beam management strategies. Future work will extend this study to urban high-rise scenarios, and will also examine the effects of UAV orientation, rotation, and antenna polarization on A2G link performance.

\section*{Acknowledgments}
This research is supported by iSEE-6G and MultiX projects under the Horizon Europe Research and Innovation program with Grant Agreement No. 101139291 and 101192521, respectively. The work of Zhuangzhuang Cui was supported by the  Research Foundation – Flanders (FWO), Senior Postdoctoral Fellowship under Grant No. 12AFN26N. The authors would also like to thank Sander Coene for his valuable assistance during the measurement campaign.

\section*{Data Availability}
The measurement dataset used in this study is publicly available and can be accessed at the following \href{https://rdr.kuleuven.be/dataset.xhtml?persistentId=doi:10.48804/MTNAEG&faces-redirect=true}{link} \cite{MTNAEG_2025}.


\balance
\bibliographystyle{IEEEtran}
\bibliography{ref}

\begin{thebibliography}{10}
\providecommand{\url}[1]{#1}
\csname url@samestyle\endcsname
\providecommand{\newblock}{\relax}
\providecommand{\bibinfo}[2]{#2}
\providecommand{\BIBentrySTDinterwordspacing}{\spaceskip=0pt\relax}
\providecommand{\BIBentryALTinterwordstretchfactor}{4}
\providecommand{\BIBentryALTinterwordspacing}{\spaceskip=\fontdimen2\font plus
\BIBentryALTinterwordstretchfactor\fontdimen3\font minus \fontdimen4\font\relax}
\providecommand{\BIBforeignlanguage}[2]{{%
\expandafter\ifx\csname l@#1\endcsname\relax
\typeout{** WARNING: IEEEtran.bst: No hyphenation pattern has been}%
\typeout{** loaded for the language `#1'. Using the pattern for}%
\typeout{** the default language instead.}%
\else
\language=\csname l@#1\endcsname
\fi
#2}}
\providecommand{\BIBdecl}{\relax}
\BIBdecl

\bibitem{10927643}
H.-J. Moon, C.-B. Chae, K.-K. Wong, and M.-S. Alouini, ``A generalized pointing error model for {FSO} links with fixed-wing {UAVs for 6G}: Analysis and trajectory optimization,'' \emph{IEEE Transactions on Wireless Communications}, vol.~24, no.~7, pp. 5723--5737, 2025.

\bibitem{siddiky2025comprehensive}
M.~N.~A. Siddiky, M.~E. Rahman, M.~S. Uzzal, and H.~D. Kabir, ``A comprehensive exploration of {6G} wireless communication technologies,'' \emph{Computers}, vol.~14, no.~1, p.~15, 2025.

\bibitem{11010845}
M.~A. Jamshed, A.~Kaushik, M.~Dajer, A.~Guidotti, F.~Parzysz, E.~Lagunas, M.~D. Renzo, S.~Chatzinotas, and O.~A. Dobre, ``Non-terrestrial networks for {6G}: Integrated, intelligent and ubiquitous connectivity,'' \emph{IEEE Communications Standards Magazine}, pp. 1--1, 2025.

\bibitem{10793277}
A.~Saboor, E.~Vinogradov, Z.~Cui, S.~Coene, W.~Joseph, and S.~Pollin, ``Elevating the future of mobility: {UAV}-enabled intelligent transportation systems,'' in \emph{2024 7th International Conference on Advanced Communication Technologies and Networking (CommNet)}, 2024, pp. 1--7.

\bibitem{guidotti2024role}
A.~Guidotti, A.~Vanelli-Coralli, M.~El~Jaafari, N.~Chuberre, J.~Puttonen, V.~Schena, G.~Rinelli, and S.~Cioni, ``Role and evolution of non-terrestrial networks toward {6G} systems,'' \emph{IEEE Access}, vol.~12, pp. 55\,945--55\,963, 2024.

\bibitem{OJCS}
A.~Saboor, E.~Vinogradov, Z.~Cui, A.~Al-Hourani, and S.~Pollin, ``A geometry-based modelling approach for the line-of-sight probability in {UAV} communications,'' \emph{IEEE Open Journal of the Communications Society}, vol.~5, pp. 364--378, 2024.

\bibitem{mohsan2022towards}
S.~A.~H. Mohsan, M.~A. Khan, F.~Noor, I.~Ullah, and M.~H. Alsharif, ``Towards the unmanned aerial vehicles ({UAVs}): A comprehensive review,'' \emph{Drones}, vol.~6, no.~6, p. 147, 2022.

\bibitem{Blacksea}
A.~Saboor, E.~Vinogradov, Z.~Cui, and S.~Pollin, ``Probability of line of sight evaluation in urban environments using {3D} simulator,'' in \emph{2023 IEEE International Black Sea Conference on Communications and Networking (BlackSeaCom)}, 2023, pp. 135--140.

\bibitem{Eucap}
A.~Saboor, Z.~Cui, E.~Vinogradov, and S.~Pollin, ``Path loss modelling for {UAV} communications in urban scenarios with random obstacles,'' in \emph{2025 19th European Conference on Antennas and Propagation (EuCAP)}, 2025, pp. 1--5.

\bibitem{gryech2024systematic}
I.~Gryech, E.~Vinogradov, A.~Saboor, P.~S. Bithas, P.~T. Mathiopoulos, and S.~Pollin, ``A systematic literature review on the role of uav-enabled communications in advancing the un’s sustainable development goals,'' \emph{Frontiers in Communications and Networks}, vol.~5, p. 1286073, 2024.

\bibitem{betti2024uav}
F.~Betti~Sorbelli, ``{UAV}-based delivery systems: A systematic review, current trends, and research challenges,'' \emph{Journal on Autonomous Transportation Systems}, vol.~1, no.~3, pp. 1--40, 2024.

\bibitem{naveen2024unlocking}
P.~Naveen, M.~P. Antony, B.~V. Ramasamy, D.~K. Sah, and R.~Maheswar, ``Unlocking the potential: How flying taxis will shape the future of transportation,'' \emph{Sustainability}, vol.~16, no.~24, p. 10795, 2024.

\bibitem{morganstanley2021_uam_tam}
{Morgan Stanley Research}, ``evtol/urban air mobility {TAM} update: A slow take-off, but sky’s the limit,'' Morgan Stanley, Technical Report, 2021.

\bibitem{agrawal2021performance}
N.~Agrawal, A.~Bansal, K.~Singh, and C.-P. Li, ``Performance evaluation of {RIS-assisted UAV-enabled} vehicular communication system with multiple non-identical interferers,'' \emph{IEEE Transactions on Intelligent Transportation Systems}, vol.~23, no.~7, pp. 9883--9894, 2021.

\bibitem{wang2021learning}
X.~Wang and M.~C. Gursoy, ``Learning-based {UAV} trajectory optimization with collision avoidance and connectivity constraints,'' \emph{IEEE Transactions on Wireless Communications}, vol.~21, no.~6, pp. 4350--4363, 2021.

\bibitem{saboor2025cash}
A.~Saboor, Z.~Cui, A.~Colpaert, E.~Vinogradov, and S.~Pollin, ``{CASH}: Context-aware smart handover for reliable {UAV} connectivity on aerial corridors,'' \emph{arXiv preprint arXiv:2508.03862}, 2025.

\bibitem{zhang2021energy}
L.~Zhang, A.~Celik, S.~Dang, and B.~Shihada, ``Energy-efficient trajectory optimization for {UAV-assisted IoT} networks,'' \emph{IEEE Transactions on Mobile Computing}, vol.~21, no.~12, pp. 4323--4337, 2021.

\bibitem{cui2022cluster}
Z.~Cui, K.~Guan, C.~Oestges, C.~Briso-Rodr{\'\i}guez, B.~Ai, and Z.~Zhong, ``Cluster-based characterization and modeling for uav air-to-ground time-varying channels,'' \emph{IEEE Transactions on Vehicular Technology}, vol.~71, no.~7, pp. 6872--6883, 2022.

\bibitem{cui2020wideband}
Z.~Cui, C.~Briso-Rodr{\'\i}guez, K.~Guan, {\.I}.~G{\"u}ven{\c{c}}, and Z.~Zhong, ``Wideband air-to-ground channel characterization for multiple propagation environments,'' \emph{IEEE Antennas and Wireless Propagation Letters}, vol.~19, no.~9, pp. 1634--1638, 2020.

\bibitem{khawaja2017uav}
W.~Khawaja, O.~Ozdemir, and I.~Guvenc, ``Uav air-to-ground channel characterization for mmwave systems,'' in \emph{2017 IEEE 86th Vehicular Technology Conference (VTC-Fall)}.\hskip 1em plus 0.5em minus 0.4em\relax IEEE, 2017, pp. 1--5.

\bibitem{tu2019low}
K.~Tu, J.~Rodr{\'\i}guez-Pi{\~n}eiro, X.~Yin, and L.~Tian, ``Low altitude air-to-ground channel modelling based on measurements in a suburban environment,'' in \emph{2019 11th International Conference on Wireless Communications and Signal Processing (WCSP)}.\hskip 1em plus 0.5em minus 0.4em\relax IEEE, 2019, pp. 1--6.

\bibitem{lv2023narrowband}
Y.~Lv, W.~Wang, and Y.~Sun, ``Narrowband {UAV} air-to-ground channel measurement and modeling in campus environment,'' in \emph{2023 17th European conference on antennas and propagation (EuCAP)}.\hskip 1em plus 0.5em minus 0.4em\relax IEEE, 2023, pp. 1--5.

\bibitem{AWPL}
A.~Saboor, Z.~Cui, E.~Vinogradov, and S.~Pollin, ``Air-to-ground channel model for pedestrian and vehicle users in general urban environments,'' \emph{IEEE Antennas and Wireless Propagation Letters}, vol.~24, no.~1, pp. 227--231, 2025.

\bibitem{saboor2025empirical}
------, ``Empirical line-of-sight probability modeling for uavs in random urban layouts,'' in \emph{2025 IEEE Wireless Communications and Networking Conference (WCNC)}.\hskip 1em plus 0.5em minus 0.4em\relax IEEE, 2025, pp. 1--6.

\bibitem{eskandari2022model}
M.~Eskandari, H.~Huang, A.~V. Savkin, and W.~Ni, ``{Model predictive control-based 3D navigation of a RIS-equipped UAV for LoS wireless communication with a ground intelligent vehicle},'' \emph{IEEE Transactions on Intelligent Vehicles}, vol.~8, no.~3, pp. 2371--2384, 2022.

\bibitem{matolak2017air}
D.~W. Matolak and R.~Sun, ``Air--ground channel characterization for unmanned aircraft systems—part iii: The suburban and near-urban environments,'' \emph{IEEE Transactions on Vehicular Technology}, vol.~66, no.~8, pp. 6607--6618, 2017.

\bibitem{xiao2025measurements}
Z.~Xiao, S.~Sun, N.~Liu, L.~Xu, and L.~Wang, ``Measurements and modeling of air-ground integrated channel in forest environment based on {OFDM} signals,'' \emph{arXiv preprint arXiv:2507.02303}, 2025.

\bibitem{colpaertMimo}
A.~Colpaert, Z.~Cui, E.~Vinogradov, and S.~Pollin, ``{3D non-stationary channel measurement and analysis for MaMIMO-UAV communications},'' \emph{IEEE Transactions on Vehicular Technology}, vol.~73, no.~5, pp. 6061--6072, 2023.

\bibitem{bai2022non}
L.~Bai, Z.~Huang, and X.~Cheng, ``A non-stationary model with time-space consistency for {6G} massive {MIMO mmWave UAV} channels,'' \emph{IEEE Transactions on Wireless Communications}, vol.~22, no.~3, pp. 2048--2064, 2022.

\bibitem{hua2025ultra}
B.~Hua, L.~Han, Q.~Zhu, C.-X. Wang, K.~Mao, J.~Bao, H.~Chang, and Z.~Tang, ``Ultra-wideband nonstationary channel modeling for {UAV}-to-ground communications,'' \emph{IEEE Transactions on Wireless Communications}, vol.~24, no.~5, pp. 4190--4204, 2025.

\bibitem{liu2021novel}
Y.~Liu, C.-X. Wang, H.~Chang, Y.~He, and J.~Bian, ``A novel non-stationary {6G UAV} channel model for maritime communications,'' \emph{IEEE Journal on Selected Areas in Communications}, vol.~39, no.~10, pp. 2992--3005, 2021.

\bibitem{bian20213d}
J.~Bian, C.-X. Wang, Y.~Liu, J.~Tian, J.~Qiao, and X.~Zheng, ``{3D non-stationary wideband UAV-to-ground MIMO channel models based on aeronautic random mobility model},'' \emph{IEEE Transactions on Vehicular Technology}, vol.~70, no.~11, pp. 11\,154--11\,168, 2021.

\bibitem{bai2022non2}
L.~Bai, Z.~Huang, and X.~Cheng, ``A non-stationary {6G UAV} channel model with 3d continuously arbitrary trajectory and self-rotation,'' \emph{IEEE Transactions on Wireless Communications}, vol.~21, no.~12, pp. 10\,592--10\,606, 2022.

\bibitem{MTNAEG_2025}
\BIBentryALTinterwordspacing
A.~Colpaert, ``{3D Massive MIMO Air-to-ground UAV CSI dataset in campus environment},'' 2025. [Online]. Available: \url{https://doi.org/10.48804/MTNAEG}
\BIBentrySTDinterwordspacing

\bibitem{willink2008wide}
T.~J. Willink, ``Wide-sense stationarity of mobile mimo radio channels,'' \emph{IEEE Transactions on Vehicular Technology}, vol.~57, no.~2, pp. 704--714, 2008.

\bibitem{cheng2022channel}
X.~Cheng, Z.~Huang, and L.~Bai, ``Channel nonstationarity and consistency for beyond 5g and 6g: A survey,'' \emph{IEEE communications surveys \& tutorials}, vol.~24, no.~3, pp. 1634--1669, 2022.

\bibitem{bultitude2002estimating}
R.~J. Bultitude, ``Estimating frequency correlation functions from propagation measurements on fading radio channels: a critical review,'' \emph{IEEE Journal on selected areas in communications}, vol.~20, no.~6, pp. 1133--1143, 2002.

\bibitem{he2015characterization}
R.~He, O.~Renaudin, V.-M. Kolmonen, K.~Haneda, Z.~Zhong, B.~Ai, and C.~Oestges, ``Characterization of quasi-stationarity regions for vehicle-to-vehicle radio channels,'' \emph{IEEE Transactions on Antennas and Propagation}, vol.~63, no.~5, pp. 2237--2251, 2015.

\bibitem{herdin2005correlation}
M.~Herdin, N.~Czink, H.~Ozcelik, and E.~Bonek, ``Correlation matrix distance, a meaningful measure for evaluation of non-stationary {MIMO} channels,'' in \emph{2005 IEEE 61st vehicular technology conference}, vol.~1.\hskip 1em plus 0.5em minus 0.4em\relax IEEE, 2005, pp. 136--140.

\bibitem{lee2006estimate}
W.~C. Lee, ``Estimate of local average power of a mobile radio signal,'' \emph{IEEE Transactions on vehicular technology}, vol.~34, no.~1, pp. 22--27, 2006.

\bibitem{Greenstein769521}
L.~Greenstein, D.~Michelson, and V.~Erceg, ``Moment-method estimation of the ricean k-factor,'' \emph{IEEE Communications Letters}, vol.~3, no.~6, pp. 175--176, 1999.

\bibitem{choi2010generation}
J.-H. Choi, S.-O. Park, T.-S. Yang, and J.-H. Byun, ``Generation of rayleigh/rician fading channels with variable rms delay by changing boundary conditions of the reverberation chamber,'' \emph{IEEE Antennas and Wireless Propagation Letters}, vol.~9, pp. 510--513, 2010.

\bibitem{Chen2017finite}
C.-M. Chen, V.~Volski, L.~Van~der Perre, G.~A.~E. Vandenbosch, and S.~Pollin, ``Finite large antenna arrays for massive {MIMO}: Characterization and system impact,'' \emph{{IEEE} Transactions on Antennas and Propagation}, vol.~65, no.~12, pp. 6712--6720, 2017.

\bibitem{massey1951ks}
\BIBentryALTinterwordspacing
F.~J. Massey, ``The kolmogorov-smirnov test for goodness of fit,'' \emph{Journal of the American Statistical Association}, vol.~46, no. 253, pp. 68--78, 1951. [Online]. Available: \url{https://doi.org/10.2307/2280095}
\BIBentrySTDinterwordspacing

\bibitem{gaertner2007characterizing}
G.~Gaertner and E.~O. Nuallain, ``Characterizing wideband signal envelope fading in urban microcells using the rice and nakagami distributions,'' \emph{IEEE transactions on vehicular technology}, vol.~56, no.~6, pp. 3621--3630, 2007.

\bibitem{cai2021characterizing}
X.~Cai, J.~Song, J.~Rodr{\'\i}guez-Pi{\~n}eiro, P.~E. Mogensen, and F.~Tufvesson, ``Characterizing the small-scale fading for low altitude uav channels,'' in \emph{The Seventeenth International Conference on Wireless and Mobile Communications}.\hskip 1em plus 0.5em minus 0.4em\relax IARIA, 2021, pp. 16--19.

\bibitem{qiu2017low}
Z.~Qiu, X.~Chu, C.~Calvo-Ramirez, C.~Briso, and X.~Yin, ``Low altitude uav air-to-ground channel measurement and modeling in semiurban environments,'' \emph{Wireless Communications and Mobile Computing}, vol. 2017, no.~1, p. 1587412, 2017.

\end{thebibliography}

\vfill

\end{document}